\documentclass[a4paper,11pt]{article}
\usepackage[margin=1in]{geometry}
\usepackage{amssymb, amsmath, amsthm, mathrsfs}
\usepackage{cite}
\usepackage{lineno}
\usepackage{tabls}
\usepackage{multirow}
\usepackage{pifont}
\usepackage{graphicx}
\usepackage{color}
\definecolor{myblue}{rgb}{0,0,1}
\definecolor{myred}{rgb}{1,0,0}

\def\qed{\hfill $\Box$\vspace{0.3cm}}
\def\pf{\noindent{\bf Proof. }}

\newtheorem{lemma}{Lemma}[section]
\newtheorem{claim}{Claim}
\newtheorem{theorem}{Theorem}
\newtheorem{remark}{Remark}[section]

\newtheorem{definition}{Definition}

\begin{document}

\title{\LARGE\bf The Component Connectivity of Alternating Group Graphs and Split-Stars}

\author{
Mei-Mei Gu$^{1,}$\thanks{This work was founded by China Postdoctoral Science Foundation (2018M631322).}
\hspace{.15in} Rong-Xia Hao$^{1,}$\thanks{This work was partially supported by the National Natural Science Foundation of China (No. 11731002) and the 111 Project of China (B16002).}
\hspace{.15in} Jou-Ming Chang$^{2,}$\thanks{Corresponding author. (This work was accomplished while the corresponding author visiting Beijing Jiaotong University, Department of Mathematics.)
} 
\\
\\
{\small $^1$ Department of Mathematics, Beijing Jiaotong University, Beijing 100044, P.R. China}\\
{\small $^2$ Institute of Information and Decision Sciences,}\\
{\small National Taipei University of Business, Taipei 10051, Taiwan}\\
{\footnotesize \emph{E-mail address}: {\tt 1212620@bjtu.edu.cn} (M.-M. Gu), {\tt rxhao@bjtu.edu.cn} (R.-X. Hao),}\\
{\footnotesize {\tt spade@ntub.edu.tw} (J.-M. Chang)}
}
\date{}
\maketitle

\begin{abstract}
\baselineskip=12pt

For an integer $\ell\geqslant 2$, the $\ell$-component connectivity of a graph $G$, denoted by $\kappa_{\ell}(G)$, is the minimum number of vertices whose removal from $G$ results in a disconnected graph with at least $\ell$ components or a graph with fewer than $\ell$ vertices. This is a natural generalization of the classical connectivity of graphs defined in term of the minimum vertex-cut and is a good measure of robustness for the graph corresponding to a network. So far, the exact values of $\ell$-connectivity are known only for a few classes of networks and small $\ell$'s. It has been pointed out in~[Component connectivity of the hypercubes, Int. J. Comput. Math. 89 (2012) 137--145] that determining $\ell$-connectivity is still unsolved for most interconnection networks, such as alternating group graphs and star graphs. In this paper, by exploring the combinatorial properties and fault-tolerance of the alternating group graphs $AG_n$ and a variation of the star graphs called split-stars $S_n^2$, we study their $\ell$-component connectivities. We obtain the following results: (i) $\kappa_3(AG_n)=4n-10$ and $\kappa_4(AG_n)=6n-16$ for $n\geqslant 4$, and $\kappa_5(AG_n)=8n-24$ for $n\geqslant 5$; (ii) $\kappa_3(S_n^2)=4n-8$, $\kappa_4(S_n^2)=6n-14$, and $\kappa_5(S_n^2)=8n-20$ for $n\geqslant 4$.

\end{abstract}
\setcounter{page}{1}
\baselineskip=14pt

%11111111111111111111111111111111111111111111
\section{Introduction}
\label{sec:intro}

An interconnection network is usually modeled as a connected graph $G=(V,E)$, where the vertex set $V(=V(G))$ represents the set of processors and the edge set $E(=E(G))$ represents the set of communication channels between processors. For a subset $S\subseteq V(G)$, the graph obtained from $G$ by removing all vertices of $S$ is denoted by $G-S$. In particular, $S$ is called a \emph{vertex-cut} of $G$ if $G-S$ is disconnected. The \emph{connectivity} of a graph $G$, denoted by $\kappa(G)$, is the cardinality of a minimum vertex-cut of $G$, or is defined to be $|V(G)|-1$ when $G$ is a complete graph. For making a more thorough study on the connectivity of a graph to assess the vulnerability of its corresponding network, a concept of generalization was first introduced by Chartrand et al.~\cite{Chart-1984}. For an integer $\ell\geqslant 2$, the \emph{generalized $\ell$-connectivity} of a graph $G$, denoted by $\kappa_\ell(G)$, is the minimum number of vertices whose removal from $G$ results in a graph with at least $\ell$ components or a graph with fewer than $\ell$ vertices. For such a generalization, a synonym was also called the \emph{general connectivity}~\cite{Sampathkumar-1984} or \emph{$\ell$-component connectivity}~\cite{Hsu-2012}. 
Since there exist diverse definitions of generalized connectivity in the literature (e.g., see~\cite{Hager-1985,Hager-1986}), hereafter we follow the use of the terminology ``$\ell$-component connectivity'' (or \emph{$\ell$-connectivity} for short) to avoid confusion.

The $\ell$-connectivity is concerned with the relevance of the cardinality of a minimum vertex-cut and the number of components caused by the vertex-cut. Accordingly, finding $\ell$-connectivity for certain interconnection networks is a good measure of robustness for such networks. So far, the exact values of $\ell$-connectivity are known only for a few classes of networks and small $\ell$'s. For example, $\ell$-connectivity is determined on 
hypercube $Q_n$ for $\ell\in[2,n+1]$ (see~\cite{Hsu-2012}) and $\ell\in[n+2,2n-4]$ (see~\cite{Zhao-2018}), 
folded hypercube $FQ_n$ for $\ell\in[2,n+2]$ (see~\cite{Zhao-2018-arXiv-0304}), 
dual cube $D_n$ for $\ell\in[2,n]$ (see~\cite{Zhao-2018-arXiv-0328}),
hierarchical cubic network $HCN(n)$ for $\ell\in[2,n+1]$ (see~\cite{Eddie-2014}), 
complete cubic network $CCN(n)$ for $\ell\in[2,n+1]$ (see~\cite{Eddie-2015}), and 
generalized exchanged hypercube $GEH(s,t)$ for $1\leqslant s\leqslant t$ and $\ell\in[2,s+1]$ (see~\cite{Eddie-2017}). 
Note that the number of vertices of graphs in the above classes is an exponent related to $n$. Also, it has been pointed out in~\cite{Hsu-2012} that determining $\ell$-connectivity is still unsolved for most interconnection networks such as star graphs $S_n$ and alternating group graphs $AG_n$. The closest results for the two classes of graph were given in~\cite{Cheng-2007,Cheng-2010}, but these are asymptotic results. Recently, Chang et al.~\cite{Chang-2018-arXiv,Chang-2018c} determined the $\ell$-connectivity of alternating group networks $AN_n$ for $\ell=3,4$. Note that the two classes of $AG_n$ and $AN_n$ are definitely different.

In this paper, we study $\ell$-connectivity of the $n$-dimensional alternating group graph $AG_n$ and the $n$-dimensional split-stars $S_n^2$ (defined later in Section~\ref{sec:Prel}), which were introduced by Jwo et al.~\cite{Jwo-1993} and Cheng et al.~\cite{Cheng-1998}, respectively, for serving as interconnection network topologies of computing systems. The two families of graphs have received much attention because they have many nice properties such as vertex-transitive, strongly hierarchical, maximally connected (i.e., the connectivity is equal to its regularity), and with a small diameter and average distance. In particular, Cheng et al.~\cite{Cheng-2002} showed that alternating group graphs and split-stars are superior to the $n$-cubes and star graphs under the comparison using an advanced vulnerability measure called toughness, which was defined in~\cite{Chvatal-1972}. For the two families of graphs, many researchers were attracted to study 
fault tolerant routing~\cite{Cheng-1999},
fault tolerant embedding~\cite{Chang-2008,Chang-2004,Tsai-2011},
matching preclusion~\cite{Bonnevilie-2011,Cheng-2008},
restricted connectivity~\cite{Cheng-2001,Gu-2017,Lin-2015,Lin-2016,Zhang-2010} and 
diagnosability~\cite{Chen-2018,Gu-2017,Hao-2013,Lin-2015-TCS,Lin-2016,Lin-2015,Tsai-2015}. 
Moreover, alternating group graphs are also edge-transitive and possess stronger and rich properties on Hamiltonicity (e.g., it has been shown to be not only pancyclic and Hamiltonian-connected~\cite{Jwo-1993} but also panconnected~\cite{Chang-2004}, panpositionable~\cite{Teng-2007} and mutually independent Hamiltonian~\cite{Su-2012}). 
The following structural property disclosed by Cheng et al.~\cite{Cheng-2010} is of particular interest and closely related to $\ell$-component connectivity. They showed that even though linearly many faulty vertices are removed in $AG_n$, the rest of the graph has still a large connected component that contains almost all the surviving vertices. Therefore, this component can be used to perform original network operations without degrading most of its capability. For more further investigations on alternating group graphs and split-stars, see also~\cite{Cheng-2000,You-2015,Zhou-2011}. 

In this paper, we determine $\ell$-component connectivity for $\ell\in\{3,4,5\}$ of the $n$-dimensional alternating group graph and $n$-dimensional split-star as follows. 

\begin{theorem}\label{thm:AGn}
$\kappa_3(AG_n)=4n-10$ and $\kappa_4(AG_n)=6n-16$ for $n\geqslant 4$, and $\kappa_5(AG_n)=8n-24$ for $n\geqslant 5$.
\end{theorem}

\begin{theorem}\label{thm:Sn2}
$\kappa_3(S_n^2)=4n-8$, $\kappa_4(S_n^2)=6n-14$, and $\kappa_5(S_n^2)=8n-20$ for $n\geqslant 4$.
\end{theorem}

%222222222222222222222222222222222222222222222222
\section{Preliminaries}
\label{sec:Prel}

For $n\geqslant 3$, let ${\mathbb Z}_n=\{1,2,\ldots,n\}$ and $p=p_1p_2\cdots p_n$ be a permutation of elements of ${\mathbb Z}_n$, where $p_i\in{\mathbb Z}_n$ is the symbol at the position $i$ in the permutation. Two symbols $p_i$ and $p_j$ are said to be a pair of \emph{inversion} of $p$ if $p_i<p_j$ and $i>j$. A permutation is an \emph{even permutation} provided it has an even number of inversions. Let ${\mathcal S}_n$ (resp., ${\mathcal A}_n$) denote the set of all permutations (resp., even permutations) over ${\mathbb Z}_n$. An operation acting on a permutation that swaps symbols at positions $i$ and $j$ and leaves all other symbols undisturbed is denoted by $\text{g}_{ij}$. The composition $\text{g}_{ij}\text{g}_{k\ell}$ means that the operation is taken by swapping symbols at positions $i$ and $j$, and then swapping symbols at positions $k$ and $\ell$. For $3\leqslant i\leqslant n$, we further define two operations, $\text{g}_i^+$ and $\text{g}_i^-$ on ${\mathcal A}_n$ by setting $\text{g}_i^+=\text{g}_{2i}\text{g}_{12}$ and $\text{g}_i^-=\text{g}_{1i}\text{g}_{12}$. Accordingly, $p\text{g}_i^+$ (resp., $p\text{g}_i^-$) is the permutation obtained from $p$ by rotating symbols at positions $1,2$ and $i$ from left to right (resp., from right to left). Taking ${\mathcal A}_5$ as an example, if $p=13425$, then $p\text{g}_4^+=21435$ and $p\text{g}_4^-=32415$. 

Recall that the \emph{Cayley graph} $Cay(X,\Omega)$ on a finite group $X$ with respect to a generating set $\Omega$ of $X$ is defined to have the vertex set $X$ and the edge set $\{(p,p\text{g})\colon\,p\in X,\text{g}\in \Omega\}$. We now formally give the definition of alternating group graphs and split-stars as follows.

\begin{definition}\label{def:AG}{\rm (see~\cite{Jwo-1993})
The \emph{$n$-dimensional alternating group graph}, denoted by $AG_n$, is a graph consisting of the vertex set $V(AG_n)={\mathcal A}_n$ and two vertices $p,q\in{\mathcal A}_n$ are adjacent if and only if $q\in\{p\text{g}_i^+,p\text{g}_i^-\}$ for some $i=3,4,\ldots,n$. That is, $AG_n=Cay({\mathcal A}_n,\Omega)$ with $\Omega=\{\text{g}_3^+,\text{g}_3^-,\text{g}_4^+,\text{g}_4^+,\ldots,\text{g}_n^+,\text{g}_n^-\}$.
}
\end{definition}

\begin{figure}[htb]
\begin{center}
\includegraphics[width=3.15in]{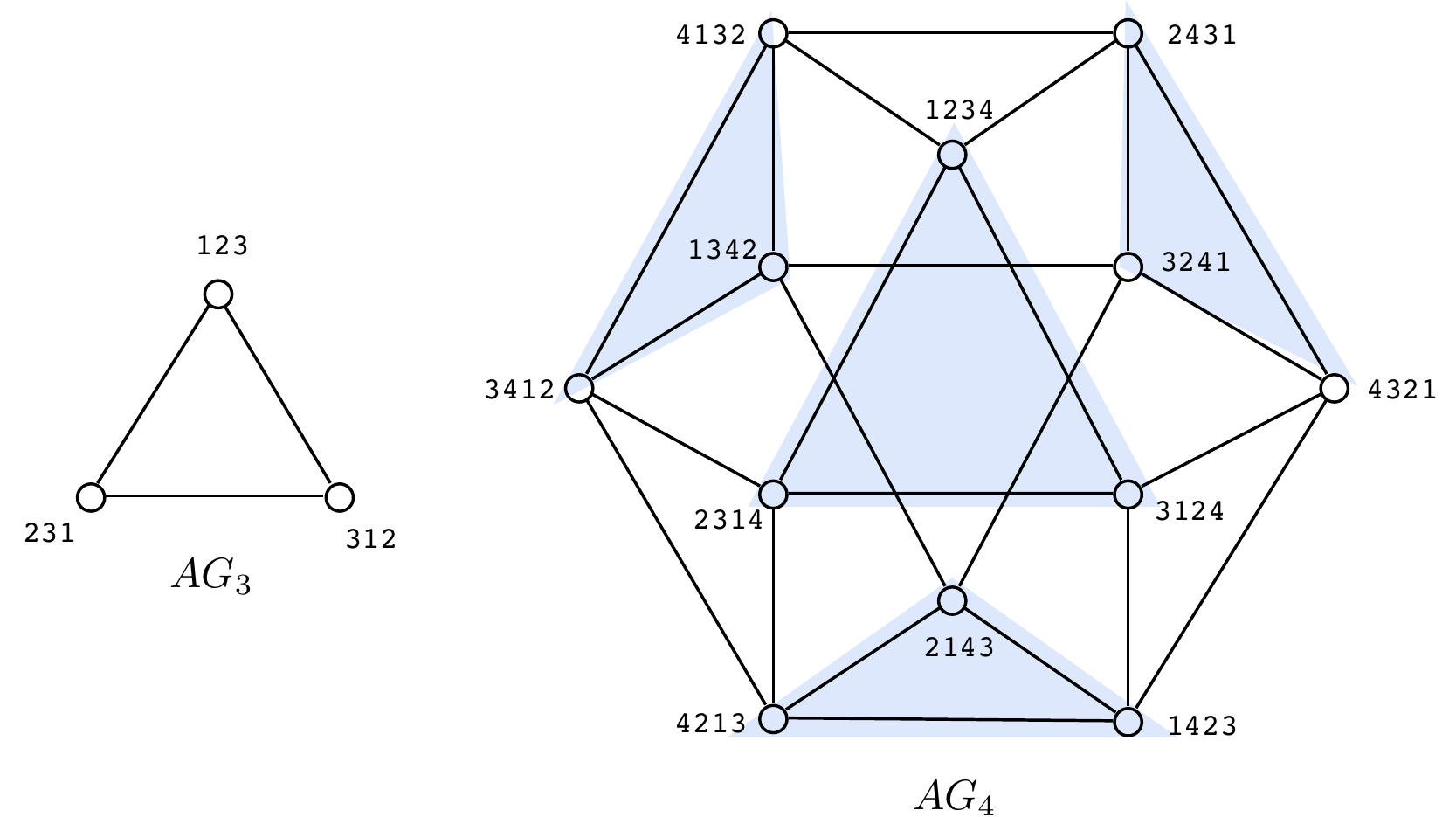}
\caption{(a) Alternating group graphs $AG_3$ and $AG_4$.} 
\label{fig:AG3-4}
\end{center}
\end{figure}

A path (resp., cycle) of length $k$ is called a $k$-path (resp., $k$-cycle). Clearly, from the above definition, $AG_3$ is isomorphic to a 3-cycle. As a Cayley graph, $AG_n$ is vertex-transitive. Also, it has been shown in~\cite{Jwo-1993} that $AG_n$ contains $n!/2$ vertices, $n!(n-2)/2$ edges, and is an edge-transitive and $(2n-4)$-regular graph with diameter $\lfloor 3n/2\rfloor-3$. It is well known that every edge-transitive graph is maximally connected, and hence $\kappa(AG_n)=2n-4$. For $n\geqslant 3$ and $i\in{\mathbb Z}_n$, let $AG_n^i$ be the subgraph of $AG_n$ induced by vertices with the rightmost symbol $i$. Like most interconnection networks, $AG_n$ can be defined recursively by a hierarchical structure. Thus, $AG_n$ is composed of $n$ disjoint copies of $AG_n^i$ for $i\in{\mathbb Z}_n$, and each $AG_n^i$ is isomorphic to $AG_{n-1}$. If a vertex $u$ belongs to a subgraph $AG_n^i$, we simply write $u\in AG_n^i$ instead of $u\in V(AG_n^i)$. An edge joining vertices in different subgraphs is an \emph{external edge}, and the two adjacent vertices are called \emph{out-neighbors} to each other. By contrast, an edge joining vertices in the same subgraph is called an \emph{internal edges}, and the two adjacent vertices are called \emph{in-neighbors} to each other. Clearly, every vertex of $AG_n$ has $2n-6$ in-neighbors and two out-neighbors. For example, Fig.~\ref{fig:AG3-4} depicts $AG_3$ and $AG_4$, where each part of shadows in $AG_4$ indicates a subgraph isomorphic to $AG_3$.

Cheng et al.~\cite{Cheng-1998} propose the Split-star networks as alternatives to the star graphs and companion graphs with the alternating group graphs.

\begin{definition}\label{def:SS}{\rm (see~\cite{Cheng-1998})
The \emph{$n$-dimensional split-star}, denoted by $S_n^2$, is a graph consisting of the vertex set $V(S_n^2)={\mathcal S}_n$ and two vertices $p,q\in{\mathcal S}_n$ are adjacent if and only if $q=p\text{g}_{12}$ or $q\in\{p\text{g}_i^+,p\text{g}_i^-\}$ for some $i=3,4,\ldots,n$. That is, $S_n^2=Cay({\mathcal S}_n,\Omega)$ with $\Omega=\{\text{g}_{12},\text{g}_3^+,\text{g}_3^-,\text{g}_4^+,\text{g}_4^+,\ldots,\text{g}_n^+,\text{g}_n^-\}$.
}
\end{definition}

\begin{figure}[htb]
\begin{center}
\includegraphics[width=4.5in]{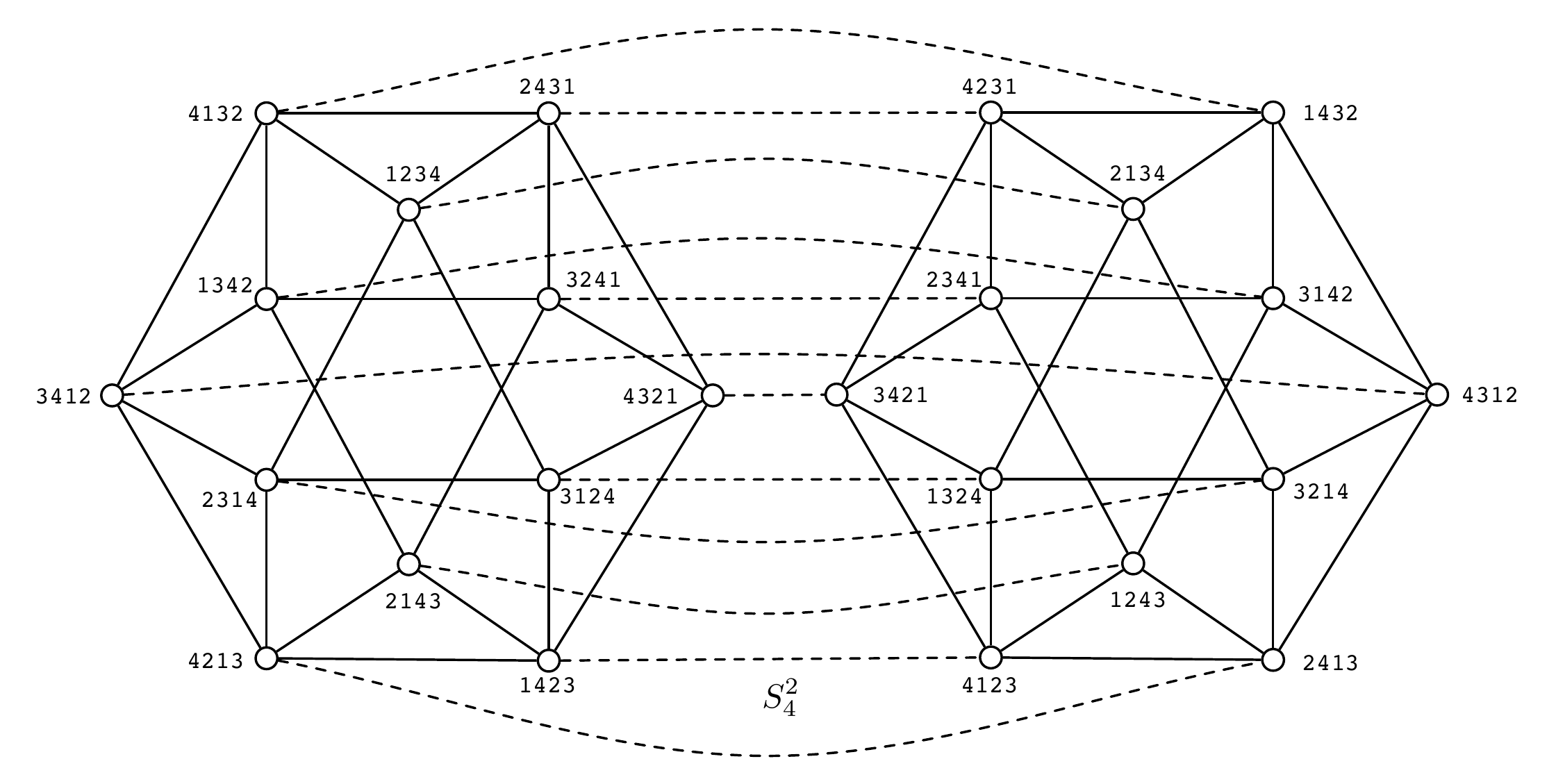}
\caption{Split-star $S_4^2$.} 
\label{fig:S4-2}
\end{center}
\end{figure}

In the above definition, the edge generated by the operation $\text{g}_{12}$ is called a \emph{$2$-exchange edge}, and others are called \emph{$3$-rotation edges}. Let $V_n^{i}$ be the set of all vertices in $S_n^2$ with the rightmost symbol $i$, i.e.,
$V_n^{i}=\{p\colon\,p=p_1p_2\cdots p_{n-1}i$, $p_j\in{\mathbb Z}_n\setminus\{i\}\ \text{for}\ 1\leqslant j\leqslant n-1\}$. Also, let $S_n^{2:i}$ denote the subgraph of $S_n^2$ induced by $V_n^{i}$. Clearly, the set $\{V_n^{i}\colon\,1\leqslant i\leqslant n\}$ forms a partition of $V(S_n^2)$ and $S_n^{2:i}$ is isomorphic to $S_{n-1}^2$. It is similar to $AG_n$ that every vertex $v\in S_n^{2:i}$ has two out-neighbors, which are joined to $v$ by external edges. Let $S_{n,E}^2$ and $S_{n,O}^2$ be subgraphs of $S_n^2$ induced by the sets of even permutations and odd permutation, respectively, in which the adjacency applied to each subgraph is precisely using the edge of $3$-rotation. Clearly, $S_{n,E}^2$ is the alternating group graph $AG_n$, and $S_{n,O}^2$ is isomorphic $S_{n,E}^2$ via a mapping $\phi(p_1p_2p_3\cdots p_n)=p_2p_1p_3\cdots p_n$ defined by $2$-exchange. Accordingly, there are $n!/2$ edges between $S_{n,E}^2$ and $S_{n,O}^2$, called \emph{matching edges}. Fig.~\ref{fig:S4-2} depicts $S_4^2$, where dashed lines indicate matching edges.

An \emph{independent set} of a graph $G$ is a subset $S\subseteq V(G)$ such that any two vertices of $S$ are nonadjacent in $G$. For $u\in V(G)$, we define $N_G(u)=\{v\in V(G):(u,v)\in E(G)\}$, i.e., the set of neighbors of $u$. Moreover, for $S\subseteq V(G)$, we define $N_G(S)=\{v\in V(G)\setminus S:\exists\ u\in S\ \text{such that}\ (u,v)\in E(G)\}$. When the graph $G$ is clear from the context, the subscript in the above notations are omitted. In what follows, we present some useful properties of $AG_n$, which will be adopted later.

\subsection{Alternating group graphs and their properties}
\label{sec:AGn}

\begin{lemma}\label{lm:basic} {\rm (see~\cite{Hao-2013})}
For $AG_n$ with $n\geqslant 4$, the following properties hold:
\begin{description}
\vspace{-0.3cm}
\item {\rm(1)} There are $(n-2)!$ external edges between any two distinct subgraphs $AG_n^i$ and $AG_n^j$ for $i,j\in{\mathbb Z}_n$ and $i\ne j$.
\vspace{-0.2cm}
\item {\rm(2)} The two out-neighbors of every vertex of $AG_n$ are contained in different subgraphs.
\vspace{-0.2cm}
\item {\rm(3)} If $u,v$ are two nonadjacent vertices of $AG_n$, then $|N(u)\cap N(v)|\leqslant 2$.
\end{description}
\end{lemma}

\begin{lemma}\label{lm:4n-11} {\rm (see~\cite{Cheng-2010})}
Let $F$ be a vertex-cut of $AG_n$ with $|F|\leqslant 4n-11$. If $n\geqslant 5$, then one of the following conditions holds:
\begin{description}
\vspace{-0.3cm}
\item {\rm(1)} $AG_n-F$ has two components, one of which is a singleton {\rm(}i.e., a trivial component{\rm)}.
\vspace{-0.2cm}
\item {\rm(2)} $AG_n-F$ has two components, one of which is an edge, say $(u,v)$. In particular, $|F|=|N(\{u,v\})|=4n-11$.
\vspace{-0.4cm}
\end{description}
Also, if $n=4$, the above description still holds except for the following two exceptions. In both cases $AG_4-F$ has two components, one of which is a $4$-cycle and the other is either a $4$-cycle {\rm(}if $|F|=4${\rm)} or a $2$-path {\rm(}if $|F|=5${\rm)}.
\end{lemma}

For example, $F=\{1234,2143,3412,4321\}$ and $F=\{1234,2143,3412,4321,2314\}$ are two exceptions of $AG_4-F$ described in Lemma~\ref{lm:4n-11}, respectively (see Fig.~\ref{fig:AG3-4}). A graph is said to be \emph{hyper-connected}~\cite{Hao-2013,Lin-2015} or \emph{tightly super-connected}~\cite{Bauer-1981} if each minimum vertex-cut creates exactly two components, one of which is a singleton. Since $\kappa(AG_4)=4$, the first exception illustrates that $AG_4$ is not hyper-connected. Here we point out a minor flaw in the literatures (e.g., see Proposition~2.4 in~\cite{Hao-2013} and Lemma~1 in~\cite{Lin-2015}), which misrepresents that $AG_4$ is hyper-connected. As a matter of fact, $AG_4$ is isomorphic to the line graph of $Q_3$ (i.e., a 3-dimensional hypercube), and the latter is contained in a list of vertex- and edge-transitive graphs without hyper-connectivity characterized by Meng~\cite{Meng-2003}. For $n\geqslant 5$, since $\kappa(AG_n)=2n-4<4n-11$, by Lemma~\ref{lm:4n-11}, $AG_n$ is hyper-connected.  

The following results are extensions of Lemma~\ref{lm:4n-11}.

\begin{lemma}\label{6n-20}{\rm(see~\cite{Cheng-2007})}
For $n\geqslant 5$, if $F$ is a vertex-cut of $AG_n$ with $|F|\leqslant 6n-20$, then one of the following conditions holds:
\begin{description}
\vspace{-0.3cm}
\item {\rm(1)} $AG_n-F$ has two components, one of which is a singleton or an edge.
\vspace{-0.2cm}
\item {\rm(2)} $AG_n-F$ has three components, two of which are singletons.
\end{description}
\end{lemma}

\begin{lemma}\label{lm:6n-19} {\rm (see~\cite{Hao-2013})}
For $n\geqslant 5$, if $F$ is a vertex-cut of $AG_n$ with $|F|\leqslant 6n-19$, then one of the following conditions holds:
\begin{description}
\vspace{-0.3cm}
\item {\rm(1)} $AG_n-F$ has two components, one of which is a singleton, an edge or a $2$-path.
\vspace{-0.2cm}
\item {\rm(2)} $AG_n-F$ has three components, two of which are singletons.
\end{description}
\end{lemma}

\begin{lemma}\label{lm:8n-29} {\rm (see~\cite{Lin-2015})}
For $n\geqslant 5$, if $F$ is a vertex-cut of $AG_n$ with $|F|\leqslant 8n-29$, then one of the following conditions holds:
\begin{description}
\vspace{-0.3cm}
\item {\rm(1)} $AG_n-F$ has two components, one of which is a singleton, an edge, a $2$-path or a $3$-cycle.
\vspace{-0.2cm}
\item {\rm(2)} $AG_n-F$ has three components, two of which are singletons or a singleton and an edge.
\vspace{-0.2cm}
\item {\rm(3)} $AG_n-F$ has four components, three of which are singletons.
\end{description}
\end{lemma}

\begin{lemma}\label{lm:AGn-singleton}
Let $S$ be an independent set of $AG_n$ for $n\geqslant 4$. Then the following assertions hold.
\begin{description}
\vspace{-0.3cm}
\item {\rm(1)} If $|S|=3$, then $|N(S)|\geqslant 6n-16$.
\vspace{-0.2cm}
\item {\rm(2)} If $|S|=4$, then $|N(S)|\geqslant 8n-24$.
\end{description}
\end{lemma}
\pf
Since $AG_n$ is vertex-transitive, one may choose the identity permutation, denoted by ${\bf e}$, as a vertex in $S$. Since $AG_n$ is $(2n-4)$-regular, if $|S|=3$ (resp., $|S|=4$) and there exists no common neighbor between any two vertices of $S$, then $|N(S)|=3(2n-4)=6n-12\geqslant 6n-16$ (resp., $|N(S)|=4(2n-4)=8n-16\geqslant 8n-24$), as required. In what follows, we assume that $N({\bf e})\cap N(S\setminus\{{\bf e}\})\ne\emptyset$ and let $N^{+}=\{{\bf e}\text{g}_i^+:i\in{\mathbb Z}_n\setminus\{1,2\}\}$ and $N^{-}=\{{\bf e}\text{g}_i^-:i\in{\mathbb Z}_n\setminus\{1,2\}\}$. Clearly, $N({\bf e})=N^+\cup N^-$ and every vertex in $N({\bf e})$ has the symbol $1,2$ or $n$ at the last position. We further define

$N^{++}\!=\!\{({\bf e}\text{g}_i^+)\text{g}_j^+:i,j\in{\mathbb Z}_n\setminus\{1,2\}\ \text{and}\ i\ne j\}$,\ 
$N^{+-}\!=\!\{({\bf e}\text{g}_i^+)\text{g}_j^-:i,j\in{\mathbb Z}_n\setminus\{1,2\}\ \text{and}\ i\ne j\}$,

$N^{-+}\!=\!\{({\bf e}\text{g}_i^-)\text{g}_j^+:i,j\in{\mathbb Z}_n\setminus\{1,2\}\ \text{and}\ i\ne j\}$,\
$N^{--}\!=\!\{({\bf e}\text{g}_i^-)\text{g}_j^-:i,j\in{\mathbb Z}_n\setminus\{1,2\}\ \text{and}\ i\ne j\}$.

\begin{figure}[htb]
\begin{center}
\includegraphics[width=\textwidth]{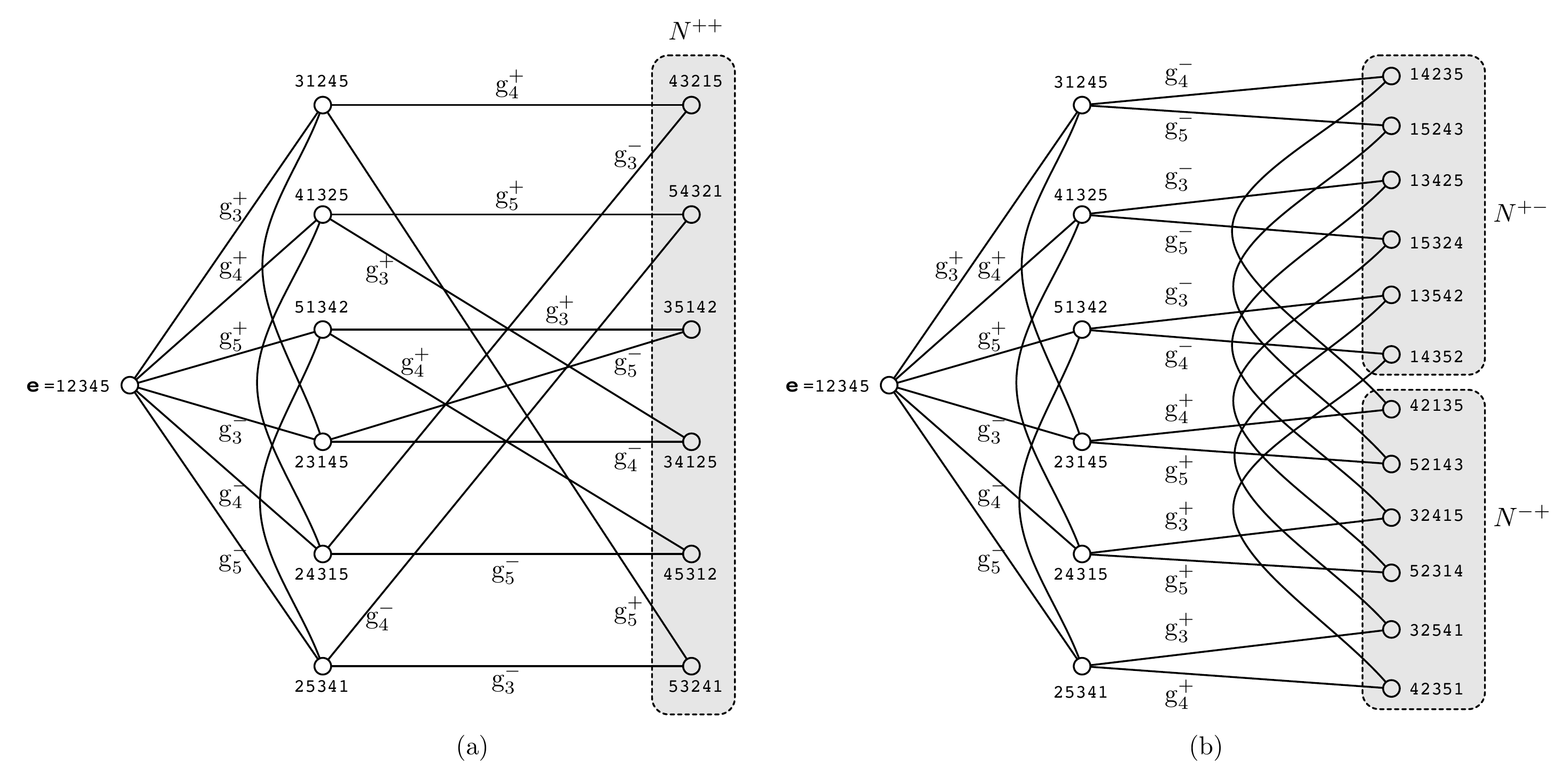}
\caption{Illustration of Lemma~\ref{lm:AGn-singleton}, where each operation $\text{g}_i^+$ or $\text{g}_i^-$ is attached to an edge between vertices (from left to right).} 
\label{fig:neighbors}
\end{center}
\end{figure}

\noindent
Since $({\bf e}\text{g}_i^+)\text{g}_j^+=({\bf e}\text{g}_j^-)\text{g}_i^-$, the two sets $N^{++}$ and $N^{--}$ are identical. If $x=({\bf e}\text{g}_i^+)\text{g}_j^+=({\bf e}\text{g}_j^-)\text{g}_i^-$, then $x$ has the symbol $j$ at the first position and symbol $i$ at the second position. In this case, we have $N({\bf e})\cap N(x)=\{{\bf e}\text{g}_i^+,{\bf e}\text{g}_j^-\}$, which meets the upper bound of Lemma~\ref{lm:basic}(3) (see Fig.~\ref{fig:neighbors}(a) for an illustration). 

\begin{claim}\label{claim-1}
For any two distinct vertices $x,y\in N^{++}$, $|N(x)\cap N(y)|\leqslant 1$. Moreover, if $z\in N(x)\cap N(y)$, then $z\in N({\bf e})$.
\end{claim}
\noindent{\em Proof of Claim}~\ref{claim-1}. 
Let $x=({\bf e}\text{g}_i^+)\text{g}_j^+$ and $y=({\bf e}\text{g}_{i'}^+)\text{g}_{j'}^+$. Consider the following situations: (i) $i=i'$ and $j\ne j'$. In this case, if there exists a common neighbor, say $z$, of $x$ and $y$, then $z=x\text{g}_j^-=(({\bf e}\text{g}_i^+)\text{g}_j^+)\text{g}_j^-=(({\bf e}\text{g}_{i'}^+)\text{g}_{j'}^+)\text{g}_{j'}^-=y\text{g}_{j'}^-$. Thus, $z={\bf e}\text{g}_i^+\in N^+$ (see, e.g., $x=43215$, $y=53241$ and $z=31245$ in Fig.~\ref{fig:neighbors}(a)); (ii) $i\ne i'$ and $j=j'$. In this case, if there exists a common neighbor, say $z$, of $x$ and $y$, then $z=x\text{g}_i^-=(({\bf e}\text{g}_i^+)\text{g}_j^+)\text{g}_i^-=(({\bf e}\text{g}_{i'}^+)\text{g}_{j'}^+)\text{g}_{i'}^-=y\text{g}_{i'}^-$. Thus, $z={\bf e}\text{g}_j^-\in N^-$ (see, e.g., $x=43215$, $y=45312$ and $z=24315$ in Fig.~\ref{fig:neighbors}(a)); (iii) $i\ne i'$ and $j\ne j'$. In this case, it is clear that $N(x)\cap N(y)=\emptyset$ (see, e.g., $x=43215$ and $y=54321$ in Fig.~\ref{fig:neighbors}(a)). This settles Claim~\ref{claim-1}.

On the other hand, the two sets $N^{+-}$ and $N^{-+}$ are not identical. Since every vertex in N({\bf e}) has two neighbors in $N^{+-}\cup N^{-+}$ and no two vertices of N({\bf e}) share a common neighbor, if $x\in N^{+-}\cup N^{-+}$, then $|N({\bf e})\cap N(x)|=1$. In fact, every vertex in $N^{+-}$ has the symbol 1 at the first position, and every vertex in $N^{-+}$ has the symbol 2 at the second position. Thus, both $N^{+-}$ and $N^{-+}$ are independent sets. Since the two symbols 1 and 2 are fixed in the first two positions for vertices in $N^{+-}$ and $N^{-+}$ respectively, every vertex in $N^{+-}$ can be adjacent to at most one vertex of $N^{-+}$, and vice versa (see Fig.~\ref{fig:neighbors}(b) for an illustration). 

\begin{claim}\label{claim-2}
For any two distinct vertices $x,y\in N^{+-}$ or $x,y\in N^{-+}$, $|N(x)\cap N(y)|\leqslant 1$. 
\end{claim}
\noindent{\em Proof of Claim}~\ref{claim-2}. 
Without loss of generality, we consider $x,y\in N^{+-}$. Let $x=({\bf e}\text{g}_i^+)\text{g}_j^-$ and $y=({\bf e}\text{g}_{i'}^+)\text{g}_{j'}^-$. Consider the following situations: (i) $i=i'$ and $j\ne j'$. In this case, if there exists a common neighbor, say $z$, of $x$ and $y$, then $z=x\text{g}_j^+=(({\bf e}\text{g}_i^+)\text{g}_j^-)\text{g}_j^+=(({\bf e}\text{g}_{i'}^+)\text{g}_{j'}^-)\text{g}_{j'}^+=y\text{g}_{j'}^+$. Thus, $z={\bf e}\text{g}_i^+\in N^+$ (see, e.g., $x=14235$, $y=15243$ and $z=31245$ in Fig.~\ref{fig:neighbors}(b)); (ii) $i\ne i'$ and $j=j'$. In this case, if there exists a common neighbor, say $z$, of $x$ and $y$, then $z=x\text{g}_i^+=(({\bf e}\text{g}_i^+)\text{g}_j^+)\text{g}_i^+=(({\bf e}\text{g}_{i'}^+)\text{g}_{j'}^+)\text{g}_{i'}^+=y\text{g}_{i'}^+$ (see, e.g., $x=14235$, $y=13425$ and $z=21435$ in Fig.~\ref{fig:neighbors}(b)); (iii) $i\ne i'$ and $j\ne j'$. In this case, it is clear that $N(x)\cap N(y)=\emptyset$ (see, e.g., $x=14235$ and $y=15324$ in Fig.~\ref{fig:neighbors}(b)). This settles Claim~\ref{claim-2}.

Note that two vertices $x\in N^{+-}$ and $y\in N^{-+}$ may have two common neighbors (see, e.g., $x=14235\in N^{+-}$ and $y=32415\in N^{-+}$ in Fig.~\ref{fig:neighbors}(b). Then $N(x)\cap N(y)=\{43215,21435\}$).

\begin{claim}\label{claim-3}
If $x\in N^{+-}\cup N^{-+}$ and $y\in N^{++}$, either $x$ and $y$ are adjacent or $|N(x)\cap N(y)|\leqslant 1$.
\end{claim}
\noindent{\em Proof of Claim}~\ref{claim-3}.
Without loss of generality, we consider $x\in N^{+-}$. Let $x=({\bf e}\text{g}_i^+)\text{g}_j^-$ and $y=({\bf e}\text{g}_{i'}^+)\text{g}_{j'}^+$. Consider the following situations: (i) $i=i'$ and $j=j'$. In this case, we have $y=({\bf e}\text{g}_{i'}^+)\text{g}_{j'}^+=(({\bf e}\text{g}_i^+)\text{g}_j^-)\text{g}_j^-=x\text{g}_j^-$, and thus $x$ and $y$ are adjacent. (ii) $i=i'$ and $j\ne j'$. In this case, if there exists a common neighbor, say $z$, of $x$ and $y$, then $z=x\text{g}_j^+=(({\bf e}\text{g}_i^+)\text{g}_j^-)\text{g}_j^+=(({\bf e}\text{g}_{i'}^+)\text{g}_{j'}^+)\text{g}_{j'}^-=y\text{g}_{j'}^-$. Thus, $z={\bf e}\text{g}_i^+\in N^+$ (see, e.g., $x=14235$, $y=53241$ and $z=31245$ in Fig.~\ref{fig:neighbors}); (iii) $i\ne i'$. In this case, it is clear that $N(x)\cap N(y)=\emptyset$. This settles Claim~\ref{claim-3}.

We are now ready to conclude the proof of the lemma. Let $v_0={\bf e}$ and $N_{i,j}=N(v_i)\cap N(v_j)$ for any tow vertices $v_i,v_j\in S$. Consider the following conditions:

For (1), let $S=\{v_0,v_1,v_2\}$. Since $N(v_0)\cap N(S\setminus\{v_0\})\ne\emptyset$, at least one vertex $v_i$ for $i=1,2$ belongs to the sets $N^{++}\cup N^{+-}\cup N^{-+}$. If $v_1,v_2\in N^{+-}\cup N^{-+}$, then $|N_{0,1}|=|N_{0,2}|=1$. Since $|N_{1,2}|\leqslant 2$ by Lemma~\ref{lm:basic}(3), it implies $|N_{0,1}\cup N_{0,2}\cup N_{1,2}|\leqslant 4$. If $v_1,v_2\in N^{++}$, then $|N_{0,1}|=|N_{0,2}|=2$. By Claim~\ref{claim-1}, we have $N_{1,2}\subset N_{0,1}\cup N_{0,2}$. Thus, $|N_{0,1}\cup N_{0,2}\cup N_{1,2}|\leqslant 4$. If $v_1\in N^{+-}\cup N^{-+}$ and $v_2\in N^{++}$ (resp., $v_2\in N^{+-}\cup N^{-+}$ and $v_1\in N^{++}$), by Claim~\ref{claim-3} either $v_1$ and $v_2$ are adjacent, which contradicts that $S$ is an independent set, or $|N_{1,2}|\leqslant 1$. Since $|N_{1,2}|\leqslant 1=|N_{0,1}|$ and $|N_{0,2}|=2$, it follows that $|N_{0,1}\cup N_{0,2}\cup N_{1,2}|\leqslant 4$. Therefore, we have $|N(S)|=3(2n-4)-|N_{0,1}\cup N_{0,2}\cup N_{1,2}|\geqslant 6n-16$ for all above situations. Also, it is clear that if $v_1\notin N^{++}\cup N^{+-}\cup N^{-+}$ or $v_2\notin N^{++}\cup N^{+-}\cup N^{-+}$, then $|N(S)|\geqslant 6n-16$. 

For (2), let $S=\{v_0,v_1,v_2,v_3\}$. Since $N(v_0)\cap N(S\setminus\{v_0\})\ne\emptyset$, at least one vertex $v_i$ for $i=1,2,3$ belongs to the sets $N^{++}\cup N^{+-}\cup N^{-+}$. Let $I={\mathbb Z}_3\cup\{0\}$ and $J=|\bigcup_{i,j\in I, i\ne j} N_{i,j}|$. 
If $v_1,v_2,v_3\in N^{++}$, then $|N_{0,i}|=2$ for $i\in{\mathbb Z}_3$ and $N_{i,j}\subset N_{0,i}\cup N_{0,j}$ for $i,j\in{\mathbb Z}_3$ and $i\ne j$ (by Claim~\ref{claim-1}). Thus, $J=6$. 
If $v_1,v_2\in N^{++}$ and $v_3\in N^{+-}\cup N^{-+}$, we have $|N_{0,1}|=|N_{0,2}|=2$, $|N_{0,3}|=1$, $N_{1,2}\subset N_{0,1}\cup N_{0,2}$ (by Claim~\ref{claim-1}), and $|N_{1,3}|,|N_{2,3}|\leqslant 1$ (by Claim~\ref{claim-3}). Thus, $J\leqslant 7$. 
If $v_1\in N^{++}$ and $v_2,v_3\in N^{+-}$ (resp., $v_1\in N^{++}$ and $v_2,v_3\in N^{-+}$), we have $|N_{0,1}|=2$, $|N_{0,2}|=|N_{0,3}|=1$, $|N_{2,3}|\leqslant 1$ (by Claim~\ref{claim-2}), and $|N_{1,2}|,|N_{1,3}|\leqslant 1$ (by Claim~\ref{claim-3}). Thus, $J\leqslant 7$. 
If $v_1\in N^{++}$, $v_2\in N^{+-}$ and $v_3\in N^{-+}$, we have $|N_{0,1}|=2$, $|N_{0,2}|=|N_{0,3}|=1$, $|N_{2,3}|\leqslant 2$ (by Lemma~\ref{lm:basic}(3)), and $|N_{1,2}|,|N_{1,3}|\leqslant 1$ (by Claim~\ref{claim-3}). Thus, $J\leqslant 8$. 
If $v_1,v_2,v_3\in N^{+-}$ (resp., $v_1,v_2,v_3\in N^{-+}$), then $|N_{0,i}|=1$ for $i\in{\mathbb Z}_3$ and $|N_{i,j}|\leqslant 1$ for $i,j\in{\mathbb Z}_3$ and $i\ne j$ (by Claim~\ref{claim-2}). Thus, $J\leqslant 6$. 
If $v_1,v_2\in N^{+-}$ and $v_3\in N^{-+}$ (resp., $v_1,v_2\in N^{-+}$ and $v_3\in N^{+-}$), we have $|N_{0,i}|=1$ for $i\in{\mathbb Z}_3$, $|N_{1,2}|\leqslant 1$ (by Claim~\ref{claim-2}), and $|N_{1,3}|,|N_{2,3}|\leqslant 2$ (by Lemma~\ref{lm:basic}(3)). Thus, $J\leqslant 8$. 
Therefore, we have $|N(S)|=4(2n-4)-J\geqslant 8n-24$ for all above situations. 
Also, if $v_i\notin N^{++}\cup N^{+-}\cup N^{-+}$ for any $i\in{\mathbb Z}_3$, by Case~1, we have $|N(S)|=|N(S\setminus\{v_i\})|+|N(v_i)|\geqslant (6n-16)+(2n-4)\geqslant 8n-24$.
\qed

Form Fig.~\ref{fig:AG3-4} it easy to check that the set $S=\{{\bf e}=1234,({\bf e}\text{g}_3^+)\text{g}_4^+=4321,({\bf e}\text{g}_4^+)\text{g}_3^+=3412\}$ (resp., $S=\{{\bf e}=1234,({\bf e}\text{g}_3^+)\text{g}_4^+=4321,({\bf e}\text{g}_4^+)\text{g}_3^+=3412,(({\bf e}\text{g}_4^+)\text{g}_3^-)\text{g}_4^+=2143\}$) is an independent set of $AG_4$ such that $N(S)=8$. Clearly, these examples show that the bounds on the assertions of Lemma~\ref{lm:AGn-singleton} are tight for $n=4$. Indeed, based on this observation, the following properties can easily be proved by induction on $n$.

\begin{remark}\label{rm:neighbor}
For $n\geqslant 4$, the following assertions hold:
\begin{description}
\vspace{-0.3cm}
\item {\rm(1)} The set $S=\{{\bf e},({\bf e}\text{g}_i^+)\text{g}_j^+,({\bf e}\text{g}_j^+)\text{g}_i^+\}$ for $i,j\in{\mathbb Z}_n\setminus\{1,2\}$ and $i\ne j$ is an independent set such that $N(S)=6n-16$.
\vspace{-0.2cm} 
\item {\rm(2)} The set $S=\{{\bf e},({\bf e}\text{g}_i^+)\text{g}_j^+,({\bf e}\text{g}_j^+)\text{g}_i^+,(({\bf e}\text{g}_j^+)\text{g}_i^-)\text{g}_j^+\}$ for $i,j\in{\mathbb Z}_n\setminus\{1,2\}$ and $i\ne j$ is an independent set such that $N(S)=8n-24$.
\end{description}
\end{remark}

%-------------------------------------------------------------------
\subsection{Split-stars and their properties}
\label{sec:Sn2}

\begin{lemma}{\rm(see~\cite{Cheng-2000,Cheng-1998,Cheng-2001})}\label{Sn2}
For $S_{n}^{2}$ with $n\geqslant 4$, the following properties hold:
\begin{enumerate}
\vspace{-0.3cm}
\item [{\rm (1)}] $S_{n}^{2}$ is $(2n-3)$-regular and $\kappa(S_{n}^{2})=2n-3$ for $n\geqslant 2$.
\vspace{-0.2cm}
\item [{\rm (2)}] The two out-neighbors of every vertex in $S_n^{2:i}$ are contained in different subgraphs and these two out-neighbors are adjacent.
For any two vertices in the same subgraph $S_n^{2:i}$, their out-neighbors in other subgraphs are different.
There are $2(n-2)!$ external edges between any two distinct subgraphs $S_n^{2:i}$ and $S_n^{2:j}$ for $i,j\in{\mathbb Z}_n$ and $i\neq j$.
\vspace{-0.2cm}
\item [{\rm (3)}] If $x,y$ are any two vertices of $S_{n}^{2}$, then
\[
 |N(x)\cap N(y)|\leqslant\left\{\begin{array}{ll}
 1 & \ \text{if $d(x,y)=1$};\\
 2 & \ \text{if $d(x,y)=2$};\\
 0 & \ \text{if $d(x,y)\geqslant 3$,}
 \end{array}\right.
\]
where $d(x,y)$ stands for the the distance {\rm(}i.e., the number of edges in a shortest path{\rm)} between $x$ and $y$ in $S_{n}^{2}$.
\end{enumerate}
\end{lemma}

\begin{lemma}{\rm(see~\cite{Cheng-2000})}\label{4n-8}
For $n\geqslant 4$, if $F$ is a vertex-cut of $S_n^2$ with $|F|\leqslant 4n-8$, then one of the following conditions holds:
\begin{description}
\vspace{-0.3cm}
\item {\rm(1)} $S_n^2-F$ has two components, one of which is a singleton.
\vspace{-0.2cm}
\item {\rm(2)} $S_n^2-F$ has two components, one of which is an edge, say $(u,v)$. If $(u,v)$ is a $2$-exchange edge, then $|F|=|N(\{u,v\})|=4n-8$; otherwise, $F=F_1\cup F_2$, where 
$F_1=N(\{u,v\})$, $|N(u)\cap N(v)|=1$, and $|F_2|\leqslant 1$.
\vspace{-0.2cm}
\item {\rm(3)} $S_n^2-F$ has three components, two of which are singletons, say $u$ and $v$. Moreover, $F=N(u)\cup N(v)$ and $|N(u)\cap N(v)|=2$, hence $|F|=4n-8$.
\end{description}
\end{lemma}

\begin{lemma}{\rm(see~\cite{Lin-2015-TCS})}\label{6n-17}
For $n\geqslant 5$, if $F$ is a vertex-cut of $S_n^2$ with $|F|\leqslant 6n-17$, then one of the following conditions holds:
\begin{description}
\vspace{-0.3cm}
\item {\rm(1)} $S_n^2-F$ has two components, one of which is a singleton, an edge or a $2$-path.
\vspace{-0.2cm}
\item {\rm(2)} $S_n^2-F$ has three components, two of which are singletons.
\end{description}
\end{lemma}

\begin{lemma}{\rm(see~\cite{Lin-2015-TCS})}\label{8n-25}
For $n\geqslant 5$, if $F$ is a vertex-cut of $S_n^2$ with $|F|\leqslant 8n-25$, then one of the following conditions holds:
\begin{description}
\vspace{-0.4cm}
\item {\rm(1)} $S_n^2-F$ has two components, one of which is a singleton, an edge, a $2$-path or a $3$-cycle.
\vspace{-0.2cm}
\item {\rm(2)} $S_n^2-F$ has three components, two of which are singletons or a singleton and an edge.
\vspace{-0.2cm}
\item {\rm(3)} $S_n^2-F$ has four components, three of which are singletons.
\end{description}
\end{lemma}

\begin{lemma}\label{singleton}
Let $S$ be an independent set of $S_n^2$ for $n\geqslant 4$. Then the following assertions hold.
\begin{description}
\vspace{-0.4cm}
\item {\rm(1)} If $|S|=2$, then $|N(S)|\geqslant 4n-8$.
\vspace{-0.2cm}
\item {\rm(2)} If $|S|=3$, then $|N(S)|\geqslant 6n-14$.
\vspace{-0.2cm}
\item {\rm(3)} If $|S|=4$, then $|N(S)|\geqslant 8n-20$.
\end{description}
\end{lemma}
\pf 
Recall that $S_n^2$ contains two copies of $AG_n$, namely $S_{n,E}^2$ and $S_{n,O}^2$. For notational convenience, we simply write $N_{S_n^2}(U)$, $N_{S_{n,E}^2}(U)$ and $N_{S_{n,O}^2}(U)$ as $N(U)$, $N_{E}(U)$ and $N_{O}(U)$ for any subset of vertices $U\subset V(S_n^2)$, respectively. Consider the following conditions:

For (1), let $S=\{v_1,v_2\}$. By Lemma~\ref{Sn2}(3), $v_1$ and $v_2$ has at most two common neighbors, $|N(S)|=|N(v_1)|+|N(v_2)|-|N(v_1)\cap N(v_2)|\geqslant 2(2n-3)-2=4n-8$.

For (2), let $S=\{v_1,v_2,v_3\}$. We consider the following cases.

\emph{Case}~2.1: Three vertices $v_1,v_2,v_3$ are contained in a common subgraph. Without loss of generality, assume $v_1,v_2,v_3\in S_{n,E}^2$. Since $S_{n,E}^2$ is isomorphic to $AG_n$, by Lemma~\ref{lm:AGn-singleton}(1), $|N_E(S)|\geqslant 6n-16$. Since each vertex of $\{v_1,v_2,v_3\}$ is joined a neighbor by a matching edge, we have $|N(S)|=|N_E(S)|+|N_O(S)|\geqslant (6n-16)+3\geqslant 6n-14$. 

\emph{Case}~2.2: Three vertices $v_1,v_2,v_3$ are distributed in two distinct subgraphs. Without loss of generality, assume $v_1,v_2\in S_{n,E}^2$ and $v_3\in S_{n,O}^2$. Since both $S_{n,E}^2$ and $S_{n,O}^2$ are isomorphic to $AG_n$, by Lemma~\ref{lm:basic}(3), $|N_E(\{v_1,v_2\})|\geqslant 2(2n-4)-2=4n-10$ and 
$|N_E(v_3)|=2n-4$. Thus, $|N(S)|\geqslant |N_E(\{v_1,v_2\})|+|N_O(v_3)|\geqslant (4n-10)+(2n-4)=6n-14$.  

For (3), let $S=\{v_1,v_2,v_3,v_4\}$. We consider the following cases.

\emph{Case}~3.1: Four vertices $v_1,v_2,v_3,v_4$ are contained in a common subgraph. Without loss of generality, assume $v_1,v_2,v_3,v_4\in S_{n,E}^2$. Since $S_{n,E}^2$ is isomorphic to $AG_n$, by Lemma~\ref{lm:AGn-singleton}(2), $|N_E(S)|\geqslant 8n-24$. Since each vertex of $\{v_1,v_2,v_3\}$ is joined a neighbor by a matching edge, we have $|N(S)|=|N_E(S)|+|N_O(S)\geqslant (8n-24)+4=8n-20$.

\emph{Case}~3.2: Four vertices $v_1,v_2,v_3,v_4$ are distributed equally in two distinct subgraphs. 
Without loss of generality, assume $v_1,v_2\in S_{n,E}^2$ and $v_3,v_4\in S_{n,O}^2$. By Lemma~\ref{lm:basic}(3), $|N_E(\{v_1,v_2\})|=|N_O(\{v_3,v_4\})|\geqslant 2(2n-4)-2=4n-10$. 
Thus, $|N(S)|\geqslant |N_E(\{v_1,v_2\})|+|N_O(\{v_3,v_4\})|\geqslant 8n-20$.

\emph{Case}~3.3: Four vertices $v_1,v_2,v_3,v_4$ are distributed nonequally in two distinct subgraphs.
Without loss of generality, assume $v_1,v_2,v_3\in S_{n,E}^2$ and $v_4\in S_{n,O}^2$. Since both $S_{n,E}^2$ and $S_{n,O}^2$ are isomorphic to $AG_n$, by Lemma~\ref{lm:AGn-singleton}(1), $|N_E(\{v_1,v_2,v_3\})|\geqslant 6n-16$ and $|N_O(v_4)|=2n-4$. Thus, $|N(S)|\geqslant |N_E(\{v_1,v_2,v_3\})|+|N_O(\{v_4\})|\geqslant 8n-20$.
\qed

%444444444444444444444444444444444444444444444444444444
\section{The $\ell$-component connectivity of $AG_n$ for $\ell\in\{3,4,5\}$}
\label{sec:main-AGn}

\begin{lemma}\label{lm:AGn-kappa-3}
For $n\geqslant 4$, $\kappa_3(AG_n)=4n-10$.
\end{lemma}
\pf
By Lemma~\ref{lm:4n-11}, if $F$ is a vertex-cut with $|F|\leqslant 4n-11$, then $AG_n-F$ has exact two components. Thus, $\kappa_3(AG_n)\geqslant 4n-10$. We now prove $\kappa_3(AG_n)\leqslant 4n-10$ as follows. For $n\geqslant 4$, since $AG_n$ is pancyclic, let $(w,x,y,z,w)$ be a 4-cycle. Also, let $F=N(\{w,y\})$. By Lemma~\ref{lm:basic}(3), we have $N(w)\cap N(y)=\{x,z\}$. Since every vertex of $AG_n$ has $2n-4$ neighbors and $w$ and $y$ share exactly two common neighbors, we have $|F|=2(2n-4)-2=4n-10$. Clearly, the removal of $F$ from $AG_n$ results in a surviving graph with a large connected component and two singletons $w$ and $y$. This attains the upper bound.
\qed

Suppose that $S$ is an independent set with the maximum cardinality in $AG_4$ and let $F=V(AG_4)\setminus S$. Obviously, $|S|=4$ (e.g., $S=\{1234,2143,3412,4321\}$) and $F$ is a vertex-cut of $AG_4$. Thus, $\kappa_4(AG_4)\leqslant 8$. From the maximality of $S$, if we choose a vertex $u\in S$, the remaining three vertices of $S$ are determined involuntary. Since $AG_4$ is vertex-transitive, $F$ is the unique vertex-cut of size 8 (up to isomorphism) in $AG_4$  such that $AG_4-F$ has four components. Thus, there is no vertex-cut $F$ with $|F|\leqslant 7$ such that $AG_4-F$ contains four components. This shows that $\kappa_4(AG_4)\geqslant 8$. As a result, we have the following lemma. 

\begin{lemma}\label{lm:kappa4-AG4}
$\kappa_4(AG_4)=8$.
\end{lemma}

We denote by $c(G)$ the number of components in a graph $G$. Hereafter, we suppose that $F$ is a vertex-cut of $AG_n$ and, for convenience, vertices in $F$ (resp., not in $F$) are called \emph{faulty vertices} (resp., \emph{fault-free vertices}). For each $i\in{\mathbb Z}_n$, let $F_i=F\cap V(AG_n^i)$, $G_i=AG_n^i-F_i$, $f_i=|F_i|$, and $c(G_i)$ be the number of components of $G_i$. Also, let $I=\{i\in{\mathbb Z}_n:G_i\ \text{is disconnected}\}$ and $J={\mathbb Z}_n\setminus I$. In addition, we adopt the following notations:
\[
F_I=\bigcup_{i\in I}F_i,\ F_J=\bigcup_{j\in J}F_j,\ AG_n^I=\bigcup_{i\in I}AG_n^i,\ \text{and}\ AG_n^J=\bigcup_{j\in J}AG_n^j.
\]

\begin{lemma}\label{lm:kappa4-AG5}
$\kappa_4(AG_5)\geqslant 14$.
\end{lemma}
\pf
Let $F$ be a vertex-cut of $AG_5$ with $|F|\leqslant 13$. Since each subgraph $AG_5^i$ is isomorphic to $AG_4$, we have $\kappa(AG_5^i)=4$. If $|I|\geqslant 4$, then $|F|\geqslant 4|I|\geqslant 16$, a contradiction. Thus, $|I|\leqslant 3$. By the definition of $J$, $G_j$ is connected for $j\in J$. If $I=\emptyset$, then $J={\mathbb Z}_5$. By Lemma~\ref{lm:basic}(1), there are $(5-2)!=6$ independent edges between $AG_5^i$ and $AG_5^j$ for $i,j\in J$ with $i\ne j$. Since $|F|\leqslant 13<3\times (5-2)!$, every $G_i$ is connected to at least two subgraphs $G_j$ and $G_k$ for $j,k\in J\setminus\{i\}$ when $I=\emptyset$. This further implies that $AG_5-F$ is connected, a contradiction. So, $1\leqslant |I|\leqslant 3$. Let $H$ be the union of components of $AG_5-F$ such that all vertices of $H$ are contained in $\bigcup_{i\in I}V(G_i)$. We claim that $AG_5^J-F_J$ is connected and $c(H)\leqslant 2$. Thus, counting together with the component that contains $AG_5^J-F_J$ as a subgraph, $AG_5-F$ contains $c(H)+1\leqslant 3$ components and the result follows. We now prove our claim by the following three cases:

{\bf Case}~1: $|I|=1$. Without loss of generality, assume $I=\{1\}$. In this case, $G_1$ is disconnected and $f_1\geqslant\kappa(AG_5^1)=4$. By Lemma~\ref{lm:basic}(1), since $|F_J|=|F|-f_1\leqslant 13-4=9<2\times (5-2)!$, every $G_i$ for $i\in J$ is connected to at least two subgraphs $G_j$ and $G_k$ for $j,k\in J\setminus\{i\}$. This further implies that $AG_5^J-F_J$ is connected. By the definition of $H$, we have $V(H)\subseteq V(G_1)$ and $H$ is not connected to $AG_5^J-F_J$. Since  by Lemma~\ref{lm:basic}(2) every vertex of $H$ has exactly two faulty out-neighbors in $F_J$, $2|V(H)|\leqslant|F_J|\leqslant 9$, which implies $|V(H)|\leqslant 4$. If $|V(H)|=4$, then $|F|-f_1=|F_J|\geqslant 2|V(H)|=8$. It follows that $f_1\leqslant |F|-8\leqslant 13-8=5=4\times 4-11$. By Lemma~\ref{lm:4n-11}, $G_1$ has two components, and thus $c(H)\leqslant c(G_1)=2$. If $|V(H)|=3$, then $c(H)\leqslant 2$. Otherwise, $H$ contains three singletons (i.e., an independent set of three vertices), and by Lemma~\ref{lm:AGn-singleton}(1), $|F|\geqslant |N_{AG_5}(V(H))|\geqslant 6\times 5-16=14$, a contradiction. Also, if $|V(H)|\leqslant 2$, it is clear that $c(H)\leqslant |V(H)|\leqslant 2$.  

{\bf Case}~2: $|I|=2$. Without loss of generality, assume $I=\{1,2\}$. Then, both $G_1$ and $G_2$ are disconnected graphs and $f_1,f_2\geqslant 4$. By Lemma~\ref{lm:basic}(1), since $|F_J|=|F|-f_1-f_2\leqslant 13-8=5<(5-2)!$, $AG_5^J-F_J$ is connected. There are two subcases as follows:

\emph{Case}~2.1: $f_1,f_2\in\{4,5\}$. For $i\in\{1,2\}$, since $f_i\leqslant 4\times 4-11$, by Lemma~\ref{lm:4n-11}, there are four situations as follows: (i) $G_i$ contains a singleton and a larger component that is connected to $AG_5^J-F_J$; (ii) $G_i$ contains an edge and a larger component that is connected to $AG_5^J-F_J$; (iii) $G_i$ contains two disjoint 4-cycles; and (iv) $G_i$ contains a 4-cycle and a 2-path. By Lemma~\ref{lm:basic}(2), every vertex of $V(G_i)$ has exactly two out-neighbors. In the latter two situations, since $|F_J|+f_j\leqslant 5+5=10<2|V(G_i)|$ where $j\in I\setminus\{i\}$, it implies that at least one component of $G_i$ must be connected to $AG_5^J-F_J$. Thus, $H$ contains at most one component of $G_i$ for $i=1,2$. This shows that $c(H)\leqslant 2$.

\emph{Case}~2.2: $f_1\geqslant 6$ (resp., $f_2\geqslant 6$). Then $|F_J|=|F|-f_1-f_2\leqslant 13-6-4=3$. By Lemma~\ref{lm:basic}(2), if a vertex $u\in F_j$ have two fault-free out-neighbors, say $u_1$ and $u_2$, in $H$, then $u_1\in V(G_1)$ and $u_2\in V(G_2)$. In this case, the vertex $u$ must be the form with a permutation $12\cdots k$ where $k\in J$. Clearly, $u_1=2k\cdots 1$ and $u_2=k1\cdots 2$. So $u_1$ and $u_2$ are adjacent in $H$. Since $|F_J|\leqslant 3$, $H$ contains at most three components, say $H_i$ for $i=1,2,3$ if they exist. Now, we show that $c(H)\leqslant 2$ by contradiction. Suppose that there exists a vertex $v_i\in V(H_i)$ for every $i\in\{1,2,3\}$. Since $H_i$ and $H_j$ are not connected in $H$ for any $i,j\in\{1,2,3\}$ with $i\ne j$, $\{v_1,v_2,v_3\}$ is an independent set of $AG_5$. Clearly, $N_{AG_5}(V(H_i))$ is a vertex-cut of $AG_5$ for each $i\in\{1,2,3\}$. Since $AG_5$ is hyper-connected, $|N_{AG_5}(V(H_i))|\geqslant \kappa(AG_5)=|N_{AG_5}(v_i)|$. By Lemma~\ref{lm:AGn-singleton}(1), $|F|\geqslant |N_{AG_5}(V(H_1)\cup V(H_2)\cup V(H_3))|\geqslant |N_{AG_5}(\{v_1,v_2,v_3\})|\geqslant 6\times 5-16=14$, a contradiction.

{\bf Case}~3: $|I|=3$. Without loss of generality, assume $I=\{1,2,3\}$. Since $|F|\leqslant 13$ and $f_i\geqslant 4$ for $i\in I$, it implies $|F_J|=|F|-f_1-f_2-f_3\leqslant 13-3\times 4=1$. By Lemma~\ref{lm:basic}(1), $AG_5^J-F_J$ is connected. Also, we have $f_i\leqslant|F|-f_j-f_k\leqslant 13-4-4=5$ for each $i\in I$ where $j,k\in I\setminus\{i\}$ with $j\ne k$. Since $f_i\in\{4,5\}$, through an argument similar to Case~2.1, we can show that $H$ contains at most one component of $G_i$, say $H_i$ if it exists, for $i=1,2,3$. If any two $H_i$ and $H_j$ are connected in $H$ for $i,j\in I$, then $c(H)\leqslant 2$. Otherwise, through an argument similar to Case~2.2 by considering
an independent set $\{v_1,v_2,v_3\}$ where $v_i\in V(H_i)$, we can show that at least one component $H_i$ for $i\in I$ does not exist. Thus, $c(H)\leqslant 2$. 
\qed

\begin{lemma}\label{lm:AGn-kappa-4}
For $n\geqslant 4$, $\kappa_4(AG_n)=6n-16$.
\end{lemma}
\pf
If $n=4$, the result is proved in Lemma~\ref{lm:kappa4-AG4}. For $n\geqslant 5$, the upper bound $\kappa_4(AG_n)\leqslant 6n-16$ can be acquired from Remark~\ref{rm:neighbor}(1) by considering the removal of $N(\{v_0,v_1,v_2\})$, where $\{v_0,v_1,v_2\}$ is an independent set of $AG_n$ and $|N(\{v_0,v_1,v_2\})|=6n-16$. Thus, the resulting graph has four components, three of which are singletons. Lemma~\ref{lm:kappa4-AG5} proves the lower bound $\kappa_4(AG_n)\geqslant 6n-16$ for $n=5$, and we now consider $n\geqslant 6$ as follows.

Let $F$ be any vertex-cut of $AG_n$ such that $|F|\leqslant 6n-17$. Lemma~\ref{lm:6n-19} shows that the removal of a vertex-cut with no more than $6n-19$ vertices in $AG_n$ results in a disconnected graph with at most three components. To complete the proof, we need to show that the same result holds when $6n-18\leqslant |F|\leqslant 6n-17$. Recall $I=\{i\in{\mathbb Z}_n:G_i\ \text{is disconnected}\}$ and $J={\mathbb Z}_n\setminus I$. By definition, $G_j$ is connected for all $j\in J$. Since $|F|\leqslant 6n-17<(n-2)!$ when $n\geqslant 6$, $AG_n^J-F_J$ remains connected for arbitrary $J$. Since $AG_n^i$ is isomorphic to $AG_{n-1}$, we have $\kappa(AG_n^i)=2n-6$. If $|I|\geqslant 4$, then $|F|\geqslant |I|\times(2n-6)\geqslant 8n-24>6n-17$, a contradiction. Also, if $I=\emptyset$, then $AG_n-F$ is connected, a contradiction. Thus, $1\leqslant |I|\leqslant 3$. Let $H$ be the union of components of $AG_n-F$ such that all vertices of $H$ are contained in $\bigcup_{i\in I}V(G_i)$. In the following, we will show that $c(H)\leqslant 2$. Thus, counting together with the component that contains $AG_n^J-F_J$ as a subgraph, $AG_n-F$ contains $c(H)+1\leqslant 3$ components. We consider the following three cases:

{\bf Case}~1: $|I|=1$. Without loss of generality, assume $I=\{1\}$. In this case, $V(H)\subseteq V(G_1)$. We analyze the number of faulty vertices of $F_J$ as follows. For $|F_J|\leqslant 7$, since every vertex of $H$ has exactly two faulty out-neighbors in $F_J$ by Lemma~\ref{lm:basic}(2), $2|V(H)|\leqslant|F_J|\leqslant 7$, which implies $|V(H)|\leqslant 3$. If $|V(H_1)|=3$, then $c(H)\leqslant 2$. Otherwise, $H_1$ contains three singletons (i.e., an independent set of three vertices), and by Lemma~\ref{lm:AGn-singleton}(1), $|F|\geqslant |N_{AG_n}(V(H))|\geqslant 6n-16$, a contradiction. Also, if $|V(H)|\leqslant 2$, it is clear that $c(H)\leqslant|V(H)|\leqslant 2$. On the other hand, we consider $|F_J|\geqslant 8$. Since $F_1$ is a vertex-cut of $AG_n^1$ and $f_1=|F|-|F_J|\leqslant (6n-17)-8=6(n-1)-19$, by Lemma~\ref{lm:6n-19}, $G_1$ contains at most three components in which the largest component is connected to $AG_n^J-F_J$. Thus, $c(G_1)\leqslant 3$ and $c(H)=c(G_1)-1\leqslant 2$. 

{\bf Case}~2: $|I|=2$. Without loss of generality, assume $I=\{1,2\}$. If $f_1\geqslant 4n-14$ or $f_2\geqslant 4n-14$, then $|F_J|=|F|-f_1-f_2\leqslant (6n-17)-(4n-14)-(2n-6)=3$. By Lemma~\ref{lm:basic}(2), every vertex of $H$ has at least one faulty out-neighbor in $F_J$. Thus, $c(H)\leqslant|V(H)|\leqslant|F_J|\leqslant 3$. If $c(H)=3$, then each component is a singleton. By Lemma~\ref{lm:AGn-singleton}(1), $|F|\geqslant N(H)\geqslant 6n-16$, a contradiction. Thus $c(H)\leqslant 2$. We now consider $f_1,f_2\leqslant 4n-15=4(n-1)-11$. For $i\in\{1,2\}$, by Lemma~\ref{lm:4n-11}, $G_i$ contains two components, one is either a singleton or an edge, and the other is a larger component connecting to $AG_n^J-F_J$. Thus, $c(G_i)=2$ for $i=1,2$ and $c(H)\leqslant c(G_1)+c(G_2)-2=2$. 

{\bf Case}~3: $|I|=3$. Without loss of generality, assume $I=\{1,2,3\}$. Since $|F|\leqslant 6n-17$ and $f_i\geqslant 2n-6$ for $i\in I$, it implies $f_i\leqslant |F|-f_j-f_k\leqslant (6n-17)-2(2n-6)=2n-5$ where $j,k\in I\setminus\{i\}$ with $j\ne k$. Since $f_i\leqslant 2n-5<4(n-1)-11$ for $n\geqslant 6$, by Lemma~\ref{lm:4n-11}, for each $i\in I$, $G_i$ contains two components, one is a singleton, say $v_i$, and the other is a larger component connecting to $AG_n^J-F_J$. If $\{v_1,v_2,v_3\}$ is an independent set of $AG_n$, by Lemma~\ref{lm:AGn-singleton}(1), $|F|\geqslant N(\{v_1,v_2,v_3\})\geqslant 6n-16$, a contradiction. Thus, at least two vertices of $v_1,v_2$ and $v_3$ are connected, which implies $c(H)\leqslant 2$.
\qed

\begin{lemma}\label{lm:kappa5-AG5}
$\kappa_5(AG_5)\geqslant 16$.
\end{lemma}
\pf Let $F$ be a vertex-cut of $AG_5$ with $|F|\leqslant 15$. Since each subgraph $AG_5^i$ is isomorphic to $AG_4$, we have $\kappa(AG_5^i)=4$. If $|I|\geqslant 4$, then $|F|\geqslant 4|I|\geqslant 16$, a contradiction. Thus, $|I|\leqslant 3$. By the definition of $J$, $G_j$ is connected for $j\in J$. If $I=\emptyset$, then $J={\mathbb Z}_5$. Through an argument similar to Lemma~\ref{lm:kappa4-AG5}, we have $AG_5-F$ is connected, a contradiction. So, $1\leqslant |I|\leqslant 3$. Let $H$ be the union of components of $AG_5-F$ such that all vertices of $H$ are contained in $\bigcup_{i\in I}V(G_i)$. We claim that $AG_5^J-F_J$ is connected and $c(H)\leqslant 3$. Thus, counting together with the component that contains $AG_5^J-F_J$ as a subgraph, $AG_5-F$ contains $c(H)+1\leqslant 4$ components and the result follows. We now prove our claim by the following three cases:

{\bf Case}~1: $|I|=1$. Without loss of generality, assume $I=\{1\}$. In this case, $G_1$ is disconnected and $f_1\geqslant\kappa(AG_5^1)=4$. By Lemma~\ref{lm:basic}(1), since $|F_J|=|F|-f_1\leqslant 15-4=11<2\times (5-2)!$, every $G_i$ for $i\in J$ is connected to at least two subgraphs $G_j$ and $G_k$ for $j,k\in J\setminus\{i\}$. This further implies that $AG_5^J-F_J$ is connected. By the definition of $H$, we have $V(H)\subseteq V(G_1)$ and $H$ is not connected to $AG_5^J-F_J$. Since by Lemma~\ref{lm:basic}(2) every vertex of $H$ has exactly two faulty out-neighbors in $F_J$, $2|V(H)|\leqslant|F_J|\leqslant 11$, which implies $|V(H)|\leqslant 5$. If $|V(H)|=5$, then $|F|-f_1=|F_J|\geqslant 2|V(H)|=10$. It follows that $f_1\leqslant |F|-10\leqslant 15-10=5=4\times 4-11$. By Lemma~\ref{lm:4n-11}, $G_1$ has two components, and thus $c(H)\leqslant c(G_1)=2$. If $|V(H)|=4$, then $c(H)\leqslant 3$. Otherwise, $H$ contains four singletons (i.e., an independent set of four vertices), and by Lemma~\ref{lm:AGn-singleton}(1), $|F|\geqslant |N_{AG_5}(V(H))|\geqslant 8\times 5-24=16$, a contradiction. Also, if $|V(H)|\leqslant 3$, it is clear that $c(H)\leqslant |V(H)|\leqslant 3$.

{\bf Case}~2: $|I|=2$. Without loss of generality, assume $I=\{1,2\}$ and $f_1\geqslant f_2$. Then, both $G_1$ and $G_2$ are disconnected graphs and $f_1\geqslant f_2\geqslant 4$. By Lemma~\ref{lm:basic}(1), since $|F_J|=|F|-f_1-f_2\leqslant 15-8=7<3(5-2)!$, $AG_5^J-F_J$ is connected. There are three subcases as follows:

\emph{Case}~2.1: $f_1,f_2\in\{4,5\}$. 
Through an argument similar to Case~2.1 in Lemma~\ref{lm:kappa4-AG5}, we know the result holds. 

\emph{Case}~2.2: $f_1\geqslant 6$ and $4\leqslant f_2\leqslant 5$. Then $|F_J|=|F|-f_1-f_2\leqslant 15-6-4=5$. Since $|F_J|\leqslant 5$, we have $|V(H)|\leqslant 5$. If $|V(H)|=5$, then $f_1=6$ and $f_2=4$. We claim $c(H)=2\leqslant 3$. For $i\in \{1,2\}$, let $H_i\subseteq H$ be the set of components such that all vertices of $H_i$ are contained in $G_i$. By Lemma~\ref{lm:kappa4-AG4} and $f_1=6<\kappa_4(AG_4)=8$, $G_1$ has at most three components and $c(H_1)\leqslant 2$. By Lemma~\ref{lm:4n-11}, $G_2$ has two components, one of which is a singleton or a four cycle and $c(H_2)=1$. It implies that $c(H)\leqslant 3$. If $|V(H)|=4$, then $c(H)\leqslant 3$. Otherwise, $H$ contains four singletons (i.e., an independent set of four vertices), and by Lemma~\ref{lm:AGn-singleton}(2), $|F|\geqslant |N_{AG_5}(V(H))|\geqslant 8\times 5-24=16$, a contradiction. Also, if $|V(H)|\leqslant 3$, it is clear that $c(H)\leqslant |V(H)|\leqslant 3$. 

\emph{Case}~2.3: $f_1,f_2\geqslant 6$. Then $|F_J|=|F|-f_1-f_2\leqslant 15-6-6=3$. This implies that $c(H)\leqslant 3$.

{\bf Case}~3: $|I|=3$. Without loss of generality, assume $I=\{1,2,3\}$ and $f_1\geqslant f_2\geqslant f_3$. Since $|F|\leqslant 15$ and $f_i\geqslant 4$ for $i\in I$, it implies $|F_J|=|F|-f_1-f_2-f_3\leqslant 15-3\times 4=3$. By Lemma~\ref{lm:basic}(1), $AG_5^J-F_J$ is connected. Also, we have $f_i\leqslant|F|-f_j-f_k\leqslant 15-4-4=7$ for each $i\in I$ where $j,k\in I\setminus\{i\}$ with $j\ne k$. There is at most one $i\in I$ such that $f_i\geqslant 6$. Otherwise, $|F|\geqslant f_1+f_2+f_3\geqslant 16>15$, a contradiction. We consider the following cases.

\emph{Case}~3.1: $4\leqslant f_3\leqslant f_2\leqslant f_1\leqslant 5$. For $i\in\{1,2,3\}$, 
by Lemma~\ref{lm:AGn-kappa-3} and $f_i\leqslant 5<\kappa_3(AG_4)=6$, $G_i$ has two components and $c(H_i)=1$. It implies that $c(H)\leqslant 3$. 

\emph{Case}~3.2: $6\leqslant f_1\leqslant 7$ and $4\leqslant f_3\leqslant f_2\leqslant 5$. 
For $i\in\{2,3\}$, by Lemma~\ref{lm:AGn-kappa-3} and $f_i\leqslant 5<\kappa_3(AG_4)=6$, $G_i$ has two components and $c(H_i)=1$. By Lemma~\ref{lm:kappa4-AG4} and $f_1\leqslant 7<\kappa_4(AG_4)=8$, $G_1$ has at most three components and $c(H_1)\leqslant 2$. Thus, $c(H)\leqslant 4$. We claim $c(H)\leqslant 3$. Suppose not and let $H_i$ for $i=1,2,3,4$ be components of $H$. Let $v_i\in V(H_i)$ for $i\in\{1,2,3,4\}$. Since $H_i$ and $H_j$ are not connected in $H$ for any $i,j\in\{1,2,3,4\}$ with $i\ne j$, $\{v_1,v_2,v_3,v_4\}$ is an independent set of $AG_5$. Clearly, $N_{AG_5}(V(H_i))$ is a vertex-cut of $AG_5$ for each $i\in\{1,2,3,4\}$. Since $AG_5$ is hyper-connected, $|N_{AG_5}(V(H_i))|\geqslant \kappa(AG_5)=|N_{AG_5}(v_i)|$. By Lemma~\ref{lm:AGn-singleton}(2), $|F|\geqslant |N_{AG_5}(V(H_1)\cup V(H_2)\cup V(H_3)\cup V(H_4))|\geqslant |N_{AG_5}(\{v_1,v_2,v_3,v_4\})|\geqslant 8\times 5-24=16$, a contradiction.
\qed

\begin{lemma}\label{lm:AGn-kappa-5}
For $n\geqslant 5$, $\kappa_5(AG_n)=8n-24$.
\end{lemma}
\pf
For $n\geqslant 5$, the upper bound $\kappa_5(AG_n)\leqslant 8n-24$ can be acquired from Remark~\ref{rm:neighbor}(2) by considering the removal of $N(\{v_0,v_1,v_2,v_3\})$, where $\{v_0,v_1,v_2,v_3\}$ is an independent set of $AG_n$ and $|N(\{v_0,v_1,v_2,v_3\})|=8n-24$. Thus, the resulting graph has five components, four of which are singletons. Lemma~\ref{lm:kappa5-AG5} proves the lower bound $\kappa_5(AG_n)\geqslant 8n-24$ for $n=5$, and we now consider $n\geqslant 6$ as follows.

Let $F$ be any vertex-cut of $AG_n$ such that $|F|\leqslant 8n-25$. Lemma~\ref{lm:8n-29} shows that the removal of a vertex-cut with no more than $8n-29$ vertices in $AG_n$ results in a disconnected graph with at most four components. To complete the proof, we need to show that the same result holds when $8n-28\leqslant |F|\leqslant 8n-25$. Recall $I=\{i\in{\mathbb Z}_n:G_i\ \text{is disconnected}\}$ and $J={\mathbb Z}_n\setminus I$. By definition, $G_j$ is connected for all $j\in J$. Since $|F|\leqslant 8n-25<(n-2)!$ when $n\geqslant 6$, $AG_n^J-F_J$ remains connected for arbitrary $J$. Since $AG_n^i$ is isomorphic to $AG_{n-1}$, we have $\kappa(AG_n^i)=2n-6$. If $|I|\geqslant 4$, then $|F|\geqslant |I|\times(2n-6)\geqslant 8n-24>8n-25$, a contradiction. Also, if $I=\emptyset$, then $AG_n-F$ is connected, a contradiction. Thus, $1\leqslant |I|\leqslant 3$. Let $H$ be the union of components of $AG_n-F$ such that all vertices of $H$ are contained in $\bigcup_{i\in I}V(G_i)$. In the following, we will show that $c(H)\leqslant 3$. Thus, counting together with the component that contains $AG_n^J-F_J$ as a subgraph, $AG_n-F$ contains $c(H)+1\leqslant 4$ components. We consider the following three cases:

{\bf Case}~1: $|I|=1$. Without loss of generality, assume $I=\{1\}$. In this case, $V(H)\subseteq V(G_1)$. We analyze the number of faulty vertices of $F_J$ as follows. 

\emph{Case}~1.1: $|F_J|\leqslant 11$. Since every vertex of $H$ has exactly two faulty out-neighbors in $F_J$ by Lemma~\ref{lm:basic}(2), $2|V(H)|\leqslant|F_J|\leqslant 11$, which implies $|V(H)|\leqslant 5$. If $|V(H)|=5$, then $c(H)\leqslant 3$. Otherwise, $H$ contains five singletons or three singletons and an edge. If $V(H)=\{v_1,v_2,v_3,v_4,v_5\}=H'\cup \{v_5\}$, where $H'=\{v_1,v_2,v_3,v_4\}$, by Lemma~\ref{lm:AGn-singleton}(2), $|N_{AG_n}(V(H))|=|N_{AG_n}(H')|+|N_{AG_n}(v_5)|-|N_{AG_n}(H')\cap N_{AG_n}(v_5)|\geqslant (8n-24)+(2n-4)-2(4\times 1)=10n-36>8n-25$ for $n\geqslant 6$, a contradiction. Now we assume $V(H)=\{v_1,v_2,v_3,u,w\}=H'\cup\{u,w\}$, where $H'=\{v_1,v_2,v_3,\}$ and $(u,w)$ is an edge. Then, by Lemma~\ref{lm:AGn-singleton}(1), $|N_{AG_n}(V(H))|=|N_{AG_n}(H')|+|N_{AG_n}(\{u,w\})|-|N_{AG_n}(H')\cap N_{AG_n}(\{u,w\})|\geqslant (6n-16)+2(2n-4)-2(3\times 2)=10n-36>8n-25$ for $n\geqslant 6$, a contradiction. If $|V(H)|=4$, then $c(H)\leqslant 3$. Otherwise, $H$ contains four singletons (i.e., an independent set of four vertices), and by Lemma~\ref{lm:AGn-singleton}(2), $|F|\geqslant |N_{AG_n}(V(H))|\geqslant 8n-24$, a contradiction. Also, if $|V(H)|\leqslant 3$, it is clear that $c(H)\leqslant|V(H)|\leqslant 3$.

\emph{Case}~1.2: $|F_J|\geqslant 12$. Since $F_1$ is a vertex-cut of $AG_n^1$ and $f_1=|F|-|F_J|\leqslant (8n-25)-12=8(n-1)-29$, by Lemma~\ref{lm:8n-29}, $G_1$ contains at most four components in which the largest component is connected to $AG_n^J-F_J$. Thus, $c(G_1)\leqslant 4$ and $c(H)=c(G_1)-1\leqslant 3$. 

{\bf Case}~2: $|I|=2$. Without loss of generality, assume $I=\{1,2\}$ and $f_1\geqslant f_2$. Since $|F|\leqslant 8n-25$ and $f_i\geqslant 2n-6$ for $i\in I$, it implies $f_i\leqslant |F|-f_j\leqslant 6n-19$ where $j\in I\setminus\{i\}$ with $j\ne i$. We consider the following subcases:

\emph{Case}~2.1: $2n-6\leqslant f_2\leqslant f_1\leqslant 4n-15=4(n-1)-11$. For $i\in\{1,2\}$, by Lemma~\ref{lm:4n-11}, $G_i$ contains two components, one is either a singleton or an edge, and the other is a larger component connecting to $AG_n^J-F_J$. Thus, $c(G_i)=2$ for $i=1,2$ and $c(H)\leqslant c(G_1)+c(G_2)-2=2$.

\emph{Case}~2.2: $2n-6\leqslant f_2\leqslant 4n-15$ and $4n-14\leqslant f_1\leqslant 6n-19$. Since $f_2\leqslant 4n-15=4(n-1)-11$, by Lemma~\ref{lm:4n-11}, $G_2$ contains two components, one is either a singleton or an edge, and the other is a larger component connecting to $AG_n^J-F_J$. Thus $c(G_2)=2$. If $4n-14\leqslant f_1\leqslant 6n-23$, by Lemma~\ref{lm:AGn-kappa-4}, $f_1<6(n-1)-16=\kappa_4(AG_{n-1})$, and thus $G_1$ contains at most three components and the largest component is connected to $AG_n^J-F_J$. Thus, $c(G_1)\leqslant 3$ and $c(H)\leqslant c(G_1)+c(G_2)-2\leqslant 3$. If $6n-22\leqslant f_1\leqslant 6n-19$, then $|F_J|=|F|-f_1-f_2\leqslant (8n-25)-(6n-22)-(2n-6)=3$. By Lemma~\ref{lm:basic}(2), every vertex of $H$ has at least one faulty out-neighbor in $F_J$. Thus, $c(H)\leqslant|V(H)|\leqslant|F_J|\leqslant 3$.

\emph{Case}~2.3: $4n-14\leqslant f_2\leqslant f_1\leqslant 6n-19$. In this case, $|F_J|=|F|-f_i-f_2\leqslant (8n-25)-2(4n-14)=3$. By Lemma~\ref{lm:basic}(2), every vertex of $H$ has at least one faulty out-neighbor in $F_J$. Thus, $c(H)\leqslant|V(H)|\leqslant|F_J|\leqslant 3$.

{\bf Case}~3: $|I|=3$. Without loss of generality, assume $I=\{1,2,3\}$ and $f_1\geqslant f_2\geqslant f_3$. Since $|F|\leqslant 8n-25$ and $f_i\geqslant 2n-6$ for $i\in I$, it implies $f_i\leqslant |F|-f_j-f_k\leqslant (8n-25)-2(2n-6)=4n-13$, where $j,k\in I\setminus\{i\}$ with $j\ne k$. We consider the following subcases:

\emph{Case}~3.1: $f_i\leqslant 4n-16<4(n-1)-11$ for each $i\in I$.
By Lemma~\ref{lm:4n-11}, $G_i$ contains two components, one is a singleton, and the other is a larger component connecting to $AG_n^J-F_J$, and thus $c(G_i)=2$. So $c(H)\leqslant c(G_1)+c(G_2)+c(G_3)-3=3\times 2-3=3$. 
 
\emph{Case}~3.2: $f_3\leqslant f_2\leqslant 4n-16< f_1\leqslant 4n-13$. In this case, each of $G_i$ for $i=2,3$ contains two components, one is a singleton, say $v_i$, and the other is a larger component connecting to $AG_n^J-F_J$. Thus $c(G_2)=c(G_3)=2$. Since $f_1\leqslant 4n-13\leqslant 6n-25=6(n-1)-19$ for $n\geqslant 6$, by Lemma~\ref{lm:6n-19}, $G_1$ contains either two components, or three components and two of which are singletons, say $v_1$ and $v_1'$. Since the largest component of $G_1$ is connected to $AG_n^J-F_J$, if $c(G_1)=2$, then $c(H)\leqslant c(G_1)+c(G_2)+c(G_3)-3=3\times 2-3=3$. On the other hand, if $\{v_1,v_1',v_2,v_3\}$ is an independent set of $AG_n$, by Lemma~\ref{lm:AGn-singleton}(2), $|F|\geqslant N(\{v_1,v_1',v_2,v_3\})\geqslant 8n-24$, a contradiction. Thus, there exists at least one of edges $(v_1,v_2)$, $(v_1,v_3)$, $(v_1',v_2)$, $(v_1',v_3)$ and $(v_2,v_3)$ in $AG_n$, which implies $c(H)\leqslant 3$.

\emph{Case}~3.3: $f_3\leqslant 4n-14\leqslant f_2\leqslant f_1\leqslant 4n-13$. Clearly, $f_3\leqslant |F|-f_1-f_2\leqslant (8n-25)-2(4n-14)=3<2n-6$ for $n\geqslant 6$, a contradiction.
 
\emph{Case}~3.4: $4n-14\leqslant f_3\leqslant f_2\leqslant f_1\leqslant 4n-3$. Clearly, $f_1+f_2+f_3\geqslant 3(4n-14)>8n-25\geqslant |F|$ when $n\geqslant 6$, a contradiction.
\qed

\noindent
{\bf Proof of Theorem~\ref{thm:AGn}.}
The result directly follows from Lemmas~\ref{lm:AGn-kappa-3}, \ref{lm:AGn-kappa-4} and \ref{lm:AGn-kappa-5}.
\qed

%555555555555555555555555555555555555555555555555555555555555555555555
\section{The $\ell$-component connectivity of $S_n^2$ for $\ell\in\{3,4,5\}$}
\label{sec:main-Sn2}

\begin{lemma}\label{lm:Sn2-kappa-3}
For $n\geqslant 4$, $\kappa_3(S_n^2)=4n-8$.
\end{lemma}
\pf
By Lemma~\ref{4n-8}, if $F$ is a vertex-cut with $|F|\leqslant 4n-9$, then $AG_n-F$ has exact two components. Thus, $\kappa_3(S_n^2)\geqslant 4n-8$. The upper bound $\kappa_3(S_n^2)\leqslant 4n-8$ can be proved using an argument similar to Lemma~\ref{lm:AGn-kappa-3} by considering that very vertex of $S_n^2$ has $2n-3$ neighbors.
\qed

\begin{lemma}\label{lm:kappa-S4-2}
$\kappa_4(S_4^2)\geqslant 10$ and $\kappa_5(S_4^2)\geqslant 12$. 
\end{lemma}
\pf
Using the notations established earlier, $S_4^2$ contains two copies of $AG_4$, say $S_{4,E}^2$ and $S_{4,O}^2$, respectively. Let $F$ be any vertex-cut of $S_4^2$. Let $F_O=F\cap V(S_{4,O}^2)$ and $F_E=F\cap V(S_{4,E}^2)$. Let $H=H_O\cup H_E$ be the union of small components of $S_n^2-F$, where $H_O$ and $H_E$ are the set of components such that their vertices are contained in $S_{n,O}^2$ and $S_{n,E}^2$, respectively.  

We first prove $\kappa_4(S_4^2)\geqslant 10$ by showing that if $|F|\leqslant 9$, then $c(H)\leqslant 3$. Note that there are $\frac{4!}{2}=12>|F|$ matching edges between $S_{4,O}^2$ and $S_{4,E}^2$. If both $S_{4,O}^2-F_O$ and $S_{4,E}^2-F_E$ are connected, then so is $S_4^2-F$, a contradiction. Next, we consider only one of $S_{4,O}^2-F_O$ and $S_{4,E}^2-F_E$ is connected. Without loss of generality, assume $S_{4,O}^2-F_O$ is connected. Then $4=\kappa(AG_4)\leqslant |F_E|\leqslant 9$. By Lemma~\ref{lm:kappa4-AG4}, if $4\leqslant |F_E|\leqslant 7<8=\kappa_4(AG_4)$, then $S_{4,E}^2-F_E$ has at most three components, and thus $c(H_E)\leqslant 2$. Since $\frac{4!}{2}=12>|F|$, the largest component of $S_{4,E}^2-F_E$ is connected to $S_{4,O}^2-F_O$, and it leads to $c(H)=c(H_E)\leqslant 2$. Also, if $8\leqslant |F_E|\leqslant 9$, then $|F_O|\leqslant 1$. Since there are $\frac{4!}{2}=12$ matching edges between $S_{4,O}^2$ and $S_{4,E}^2$, every component of size at least $2$ in $S_{4,E}^2-F_E$ is part of the component in $S_4^2-F$ containing $S_{n,O}^2-F_O$, and at most one vertex in $S_{4,E}^2-F_E$ is not part of this component containing $S_{4,O}^2-F_O$. Thus, $|V(H_E)|\leqslant 1$ and $c(H)\leqslant |V(H_E)|\leqslant 1$. We now consider both $S_{4,O}^2-F_O$ and $S_{4,E}^2-F_E$ are disconnected. Without loss of generality, assume $|F_O|\geqslant |F_E|\geqslant 4$. Since $|F|\leqslant 9$, it implies $4\leqslant |F_E|\leqslant |F_O|\leqslant 5$. By Lemma~\ref{lm:4n-11}, each of $S_{4,O}^2-F_O$ and $S_{4,E}^2-F_E$ has two components. Thus, $c(H_O)=c(H_E)=1$. Since the largest component of $S_{4,E}^2-F_E$ is connected to $S_{4,O}^2-F_O$, it leads to $c(H)\leqslant c(H_O)+c(H_E)=2$. 

Next, we prove $\kappa_5(S_4^2)\geqslant 12$ by showing that if $|F|\leqslant 11$, then $c(H)\leqslant 4$. Note that there are $\frac{4!}{2}=12>|F|$ matching edges between $S_{4,O}^2$ and $S_{4,E}^2$. If both $S_{4,O}^2-F_O$ and $S_{4,E}^2-F_E$ are connected, then so is $S_4^2-F$, a contradiction. Next, we consider only one of $S_{4,O}^2-F_O$ and $S_{4,E}^2-F_E$ is connected. Without loss of generality, assume $S_{4,O}^2-F_O$ is connected. Then $4=\kappa(AG_4)\leqslant |F_E|\leqslant 11$. If $4\leqslant |F_E|\leqslant 7<8=\kappa_4(AG_4)$, we can show that $c(H)\leqslant 2$ through a similar discussion as above. So we assume $8\leqslant |F_E|\leqslant 11$, and this implies $|F_O|\leqslant 3$. Since there are $\frac{4!}{2}=12$ matching edges between $S_{4,O}^2$ and $S_{4,E}^2$, every component of size at least $4$ in $S_{4,E}^2-F_E$ is part of the component in $S_4^2-F$ containing $S_{n,O}^2-F_O$, and at most three vertex in $S_{4,E}^2-F_E$ is not part of this component containing $S_{4,O}^2-F_O$. Thus, $|V(H_E)|\leqslant 3$ and $c(H)\leqslant |V(H_E)|\leqslant 3$. We now consider both $S_{4,O}^2-F_O$ and $S_{4,E}^2-F_E$ are disconnected. Without loss of generality, assume $|F_O|\geqslant |F_E|\geqslant 4$. Since $|F|\leqslant 11$, it implies $4\leqslant |F_E|\leqslant |F_O|\leqslant 7$ and at most one $i\in \{E,O\}$ such that $|F_i|\geqslant 6$. If $4\leqslant |F_E|\leqslant |F_O|\leqslant 5$, we can show that $c(H)\leqslant 2$ through a similar discussion as above. Finally, we consider $6\leqslant |F_O|\leqslant 7$ and $4\leqslant |F_E|\leqslant 5$. By Lemma~\ref{lm:kappa4-AG4}, $6\leqslant |F_O|\leqslant 7<8=\kappa_4(AG_4)$ implies that $S_{4,O}^2-F_O$ has at most three components and $c(H_O)\leqslant 2$. Also, by Lemma~\ref{lm:4n-11}, $4\leqslant |F_E|\leqslant 5$ implies that $S_{4,E}^2-F_E$ has two components and $c(H_E)=1$. Since the largest component of $S_{4,E}^2-F_E$ is connected to the largest component of $S_{4,O}^2-F_O$, we have $c(H)\leqslant c(H_E)+c(H_O)\leqslant 3$.
\qed

\begin{lemma}\label{lm:Sn2-kappa-4}
For $n\geqslant 4$, $\kappa_4(S_n^2)=6n-14$.
\end{lemma}
\pf
For $n\geqslant 4$, the upper bound $\kappa_4(S_n^2)\leqslant 6n-14$ can be acquired from Lemma~\ref{singleton}(2) by considering the removal of $N_{S_n^2}(\{v_1,v_2,v_3\})$ where $\{v_1,v_2,v_3\}$ is an independent set of $S_n^2$, and thus the resulting graph has four components, three of which are singletons. By Lemma~\ref{lm:kappa-S4-2}, we know $\kappa_4(S_4^2)\geqslant 10=6\times 4-14$. So we prove the lower bound $\kappa_4(S_n^2)\geqslant 6n-14$ for $n\geqslant 5$ as follows. Recall that $S_n^2$ contains two copies of $AG_n$, say $S_{n,E}^2$ and $S_{n,O}^2$, respectively. Let $F$ be any vertex-cut of $S_n^2$ such that $|F|\leqslant 6n-15$. Lemma~\ref{6n-17} shows that the removal of a vertex-cut with no more than $6n-17$ vertices in $S_n^2$ results in a disconnected graph with at most three components. To complete the proof, we need to show that the same result holds when $6n-16\leqslant |F|\leqslant 6n-15$. 

Let $F_O=F\cap V(S_{n,O}^2)$ and $F_E=F\cap V(S_{n,E}^2)$. Let $H=H_O\cup H_E$ be the union of small components of $S_n^2-F$, where $H_O$ and $H_E$ are the set of components such that their vertices are contained in $S_{n,O}^2$ and $S_{n,E}^2$, respectively. Without loss of generality, assume $|F_O|\geqslant |F_E|$. Since $2(4n-11)>6n-15$ for $n\geqslant 5$, we consider the following two cases.

{\bf Case}~1: $|F_E|\leqslant |F_O|\leqslant 4n-12$. By Lemma~\ref{lm:4n-11}, $S_{n,O}^2-F_O$ (resp., $S_{n,E}^2-F_E$) either is connected or has two components, one of which is a singleton. Let $B_O$ (resp., $B_E$) be the largest component of $S_{n,O}^2-F_O$ (resp., $S_{n,E}^2-F_E$). Since $\frac{n!}{2}-(6n-15)-2>0$ for $n\geqslant 5$, $B_O$ and $B_E$ belong to the same component in $S_n^2-F$. Note that $F$ is a vertex-cut of $S_n^2$, the singletons in $S_{n,O}^2-F_O$ and $S_{n,E}^2-F_E$ can remain singleton or for two of them to form an edge in $S_n^2-F$. Thus, $S_n^2-F$ has at most three components, i.e. $c(H)\leqslant 2$. The result holds.

{\bf Case}~2: $4n-11\leqslant |F_O|\leqslant 6n-15$. It implies that $|F_E|\leqslant (6n-15)-(4n-11)\leqslant 2n-4$. Note that $S_{n,E}^2$ is isomorphic to $AG_n$ and $2n-4\leqslant 4n-12$ for $n\geqslant 5$, by Lemma~\ref{lm:4n-11}, so $S_{n,E}^2-F_E$ either is connected or has two components, one of which is a singleton. Thus $V(H_E)\leqslant 1$ and $c(H_E)\leqslant 1$. If $S_{n,O}^2-F_O$ is connected, note that $\frac{n!}{2}-(6n-15)-1>0$ for $n\geqslant 5$, then $S_n^2-F$ has two components, one of which is a singleton. The result holds in this case. In the following, we assume that $S_{n,O}^2-F_O$ is disconnected, and consider the following cases:

\emph{Case}~2.1: $6n-18\leqslant |F_O|\leqslant 6n-15$. It implies $|F_E|\leqslant(6n-15)-(6n-18)=3$, and thus $S_{n,E}^2-F_E$ is connected. Note that there are $\frac{n!}{2}$ matching edges between $S_{n,O}^2$ and $S_{n,E}^2$. Since $|F_E|\leqslant 3$, every component of size at least $4$ in $S_{n,O}^2-F_O$ is part of the component in $S_n^2-F$ containing $S_{n,E}^2-F_E$, and at most three vertices in $S_{n,O}^2-F_O$ are not part of this component containing $S_{n,E}^2-F_E$. Thus, $|V(H_O)|\leqslant 3$ and $|V(H)|=|V(H_O)|+|V(H_E)|\leqslant 4$. If $|V(H)|=4$, then $c(H)\leqslant 2$. Otherwise, $H$ contains four singletons or two singletons and an edge. If $H$ contains four singletons, by Lemma~\ref{singleton}, $|N_{S_n^2}(H)|\geqslant 8n-20>6n-15$ for $n\geqslant 5$, a contradiction. Now we assume that $V(H)=\{v_1,v_2,u,w\}=H'\cup\{u,w\}$, where $H'=\{v_1,v_2\}$ and $(u,w)$ is an edge. Then, by Lemma~\ref{singleton}(1) and Lemma~\ref{Sn2}(3), $|N_{S_n^2}(V(H))|=|N_{S_n^2}(H')|+|N_{S_n^2}(\{u,v\})|-|N_{S_n^2}(H')\cap N_{S_n^2}(\{u,v\})|\geqslant (4n-8)+2(2n-3)-2\times 3=8n-20>6n-15$ for $n\geqslant 5$, a contradiction. If $|V(H)|=3$, then $c(H)\leqslant 2$. Otherwise, $H$ contains three singletons, and by Lemma~\ref{singleton}(2), $|F|\geqslant |N_{S_n^2}(V(H))|\geqslant 6n-14$, a contradiction. Also, if $|V(H)|\leqslant 2$, it is clear that $c(H)\leqslant|V(H)|\leqslant 2$.

\emph{Case}~2.2: $4n-11\leqslant |F_O|\leqslant 6n-19$. It implies $|F_E|\leqslant(6n-15)-(4n-11)=2n-4$, and thus $S_{n,E}^2-F_E$ is connected. By Lemma~\ref{lm:6n-19}, $S_{n,O}^2-F_O$ either has two components, one of which is a singleton, an edge or a $2$-path, or has three components, two of which are singletons. Let $C$ be the largest component of $S_{n,O}^2-F_O$. Since $\frac{n!}{2}-(6n-15)-3>0$ for $n\geqslant 5$, $C$ is part of the component in $S_n^2-F$ containing $S_{n,E}^2-F_E$. Thus, $|V(H_O)|\leqslant 3$ and $|V(H)|=|V(H_O)|+|V(H_E)|\leqslant 4$. Then, through a similar argument in the above case, we can show that $c(H)\leqslant 2$. 
\qed

\begin{lemma}\label{lm:Sn2-kappa-5}
For $n\geqslant 4$, $\kappa_5(S_n^2)=8n-20$.
\end{lemma}
\pf
For $n\geqslant 4$, the upper bound $\kappa_5(S_n^2)\leqslant 8n-20$ can be acquired from Lemma~\ref{singleton} by considering the removal of $N_{S_n^2}(\{v_1,v_2,v_3,v_4\})$ where $\{v_1,v_2,v_3,v_4\}$ is an independent set of $S_n^2$, and thus the resulting graph has five components, four of which are singletons. By Lemma~\ref{lm:kappa-S4-2}, we know $\kappa_5(S_4^2)\geqslant 12=8\times 4-20$. So we prove the lower bound $\kappa_5(S_n^2)\geqslant 8n-20$ for $n\geqslant 5$ as follows. Let $F$ be any vertex-cut of $S_n^2$ such that $|F|\leqslant 8n-21$. Lemma~\ref{8n-25} shows that the removal of a vertex-cut with no more than $8n-25$ vertices in $S_n^2$ results in a disconnected graph with at most four components. To complete the proof, we need to show that the same result holds when $8n-24\leqslant |F|\leqslant 8n-19$.

Let $F_O=F\cap V(S_{n,O}^2)$ and $F_E=F\cap V(S_{n,E}^2)$. Let $H=H_O\cup H_E$ be the union of small components of $S_n^2-F$, where $H_O$ and $H_E$ are the set of components such that their vertices are contained in $S_{n,O}^2$ and $S_{n,E}^2$, respectively. Without loss of generality, assume $|F_O|\geqslant |F_E|$. Since $2(6n-19)>8n-21$ for $n\geqslant 5$, we consider the following cases. 

{\bf Case}~1: $|F_E|\leqslant |F_O|\leqslant 4n-12$. By Lemma~\ref{lm:4n-11}, $S_{n,O}^2-F_O$ (resp., $S_{n,E}^2-F_E$) either is connected or has two components, one of which is a singleton. Since $\frac{n!}{2}-(8n-21)-2>0$ for $n\geqslant 5$, a proof similar to Case~1 in Lemma~\ref{lm:Sn2-kappa-4} can show that $c(H)\leqslant 2$.

{\bf Case}~2: $4n-11\leqslant |F_E|\leqslant |F_O|\leqslant 6n-20$. By Lemma~\ref{6n-20}, $S_{n,O}^2-F_O$ (resp., $S_{n,E}^2-F_E$) has at most three components, and $|V(H_O)|\leqslant 2$ (resp., $|V(H_E)|\leqslant 2$). Thus, $|V(H)|\leqslant 4$. Since $\frac{n!}{2}-(8n-21)-4>0$ for $n\geqslant 5$, the largest component of $S_{n,O}^2-F_O$ is connected to the largest component of $S_{n,E}^2-F_E$. If $|V(H)|=4$, then $c(H)\leqslant 3$. Otherwise, by Lemma~\ref{singleton}(3), $|N_{S_n^2}(H)|\geqslant 8n-20>8n-21$ for $n\geqslant 5$, a contradiction. Also, if $|V(H)|\leqslant 3$, it is clear that $c(H)\leqslant|V(H)|\leqslant 3$.

{\bf Case}~3: $6n-19\leqslant |F_O|\leqslant 8n-21$. In this case, $|F_E|\leqslant 8n-21-(6n-19)=2n-2\leqslant 4n-12$. By Lemma~\ref{lm:4n-11}, $S_{n,E}^2-F_E$ has at most two components and $|V(H_E)|\leqslant 1$. Thus $c(H_E)\leqslant 1$. If $S_{n,O}^2-F_O$ is connected, note that $\frac{n!}{2}-(8n-21)-1>0$ for $n\geqslant 5$, then $S_n^2-F$ has two components, one of which is a singleton. The result holds in this case. In the following, we assume that $S_{n,O}^2-F_O$ is disconnected, and consider the following cases:

\emph{Case}~3.1: $8n-24\leqslant |F_O|\leqslant 8n-21$. It implies $|F_E|\leqslant(8n-21)-(8n-24)=3$, and thus $S_{n,E}^2-F_E$ is connected. Then a proof similar to Case~2.1 in Lemma~\ref{lm:Sn2-kappa-4} can show that $|V(H)|\leqslant 4$. If $|V(H_1)|=4$, then $c(H)\leqslant 3$. Otherwise, $H_1$ contains four singletons, and by Lemma~\ref{singleton}, $|F|\geqslant |N_{S_n^2}(V(H))|\geqslant 8n-20$, a contradiction. Also, if $|V(H)|\leqslant 3$, it is clear that $c(H)\leqslant|V(H)|\leqslant 3$.

\emph{Case}~3.2: $6n-19\leqslant |F_O|\leqslant 8n-25$. By Lemma~\ref{lm:AGn-kappa-5}, $\kappa_5(AG_n)=8n-24$. Since $6n-19\leqslant |F_O|\leqslant 8n-25<8n-24$, $S_{n,O}^2-F_O$ has at most four components and $c(H_O)\leqslant 3$. As before, the largest component of $S_{n,O}^2-F_O$ is connected to the largest component of $S_{n,E}^2-F_E$. It implies that $c(H)\leqslant c(H_O)+c(H_E)\leqslant 4$.
\qed

\noindent
{\bf Proof of Theorem~\ref{thm:Sn2}.}
The result directly follows from Lemmas~\ref{lm:Sn2-kappa-3}, \ref{lm:Sn2-kappa-4} and \ref{lm:Sn2-kappa-5}.
\qed

%666666666666666666666666666666666666666666666666666666666666666666666
\section{Concluding remarks}
\label{sec:Conclusion}

In this paper, we study the $\ell$-component connectivity of alternating group graphs and split-stars. For alternating group graphs, we obtain the results: $\kappa_3(AG_n)=4n-10$ and $\kappa_4(AG_n)=6n-16$ for $n\geqslant 4$, and $\kappa_5(AG_n)=8n-24$ for $n\geqslant 5$. For split-stars, we obtain the results: $\kappa_3(S_n^2)=4n-8$ for $n\geqslant 4$, and $\kappa_4(S_n^2)=6n-14$ and $\kappa_5(S_n^2)=8n-20$ for $n\geqslant 5$. So far the problem of determining $\kappa_{\ell}(AG_n)$ and $\kappa_{\ell}(S_n^2)$ for $\ell\geqslant 6$ are still open. 

F\`abrega and Fiol~\cite{Fabrega-2017} introduced another evaluation of the reliability for interconnection networks. Given a graph $G$ and a nonnegative integer $h$, the \emph{$h$-extra connectivity} of $G$, denoted by $\kappa^{(h)}(G)$, is the cardinality of a minimum vertex-cut $S$ of $G$, if it exists, such that each component of $G-S$ has at least $h+1$ vertices. In fact, the extra connectivity plays an important indicator of a network's ability for diagnosis and fault tolerance~\cite{Gu-2017,Hao-2016,Lin-2016,Lin-2015}. Currently, the known results of $h$-extra connectivity for alternating group graphs and split-stars were proposed in \cite{Lin-2015} and \cite{Lin-2016}, respectively. The following table compares the two types of connectivities for alternating group graphs and split-stars. From this table, it seems that $\kappa_\ell(G)$ and $\kappa^{(\ell-2)}(G)$ have strongly close relationship for a network $G$. Based on the current resultst $\kappa_{\ell}(G)>\kappa^{(\ell-2)}(G)$ for $G\in\{AG_n,S_n^2\}$ and $\ell\in\{3,4,5\}$, finding $\kappa_{\ell}(G)$ needs more analyses than that of $\kappa^{(\ell-2)}(G)$. An interesting question is that does the relation always hold for larger $\ell$?

%similar results for $n$-dimensional alternating group networks $AN_n$ are known as follows: $\kappa_3(AN_n)=2n-3$ and $\kappa_4(AN_n)=3n-6$ for $n\geqslant 4$~\cite{Chang-2018-arXiv,Chang-2018c} and $\kappa^{(1)}(AN_n)=2n-5$ and $\kappa^{(2)}(AN_n)=3n-9$ for $n\geqslant 4$~\cite{Zhou-2009c}. 

\begin{table}[htp]\small
\begin{center}
\begin{tabular}{|c||l|c||l|c|}\hline
Graph classes & $h$-extra connectivity & Ref. & $\ell$-component connectivity & Ref. \\ \hline
        & $\kappa^{(1)}(AG_n)=4n-11$ for $n\geqslant 5$ &  & $\kappa_3(AG_n)=4n-10$ for $n\geqslant 4$ & \\
$AG_n$  & $\kappa^{(2)}(AG_n)=6n-19$ for $n\geqslant 5$ & \cite{Lin-2015} & $\kappa_4(AG_n)=6n-16$ for $n\geqslant 4$ & \\
        & $\kappa^{(3)}(AG_n)=8n-28$ for $n\geqslant 5$ &  & $\kappa_5(AG_n)=8n-24$ for $n\geqslant 5$ & this\\ \cline{1-4}
        & $\kappa^{(1)}(S_n^2)=4n-9$  for $n\geqslant 4$ &  & $\kappa_3(S_n^2)=4n-8$ for $n\geqslant 4$ & paper \\
$S_n^2$ & $\kappa^{(2)}(S_n^2)=6n-16$ for $n\geqslant 4$ & \cite{Lin-2016} & $\kappa_4(S_n^2)=6n-14$ for $n\geqslant 4$& \\
        & $\kappa^{(3)}(S_n^2)=8n-24$ for $n\geqslant 4$ &  & $\kappa_5(S_n^2)=8n-20$ for $n\geqslant 4$ & \\ \hline
\end{tabular}
\end{center}
\label{tbl:Compares}
\end{table}%

\end{document}